\documentclass[fleqn,usenatbib]{mnras}

\usepackage{newtxtext,newtxmath}
\usepackage[T1]{fontenc}

\DeclareRobustCommand{\VAN}[3]{#2}
\let\VANthebibliography\thebibliography
\def\thebibliography{\DeclareRobustCommand{\VAN}[3]{##3}\VANthebibliography}

\usepackage{graphicx}	% Including figure files
\usepackage{amsmath}	% Advanced maths commands
\usepackage{multirow}
\usepackage{booktabs}
\usepackage{soul}
\usepackage{orcidlink}
\soulregister{\ref}7
\usepackage{color, xcolor}
\usepackage{amssymb}	% Extra maths symbols

\title[VZOP Algorithm]{Detecting Cosmic\,21 cm Global Signal Using an Improved Polynomial Fitting Algorithm}

\author[T Liu et al.]{
	Tianyang Liu\,\orcidlink{0009-0001-0228-9130}\,$^{1,2}$,
	Junhua Gu\,\orcidlink{0000-0001-9765-6521}\,$^{3}$\thanks{E-mail: jhgu@nao.cas.cn (JG)},
	Quan Guo\,\orcidlink{0000-0003-2858-5090}\,$^{1}$\thanks{E-mail: guoquan@shao.ac.cn (QG)},
	Huanyuan Shan\,\orcidlink{0000-0001-8534-837X}\,$^{1}$,
	Qian Zheng$^{1}$,
	Jingying Wang\,\orcidlink{0000-0002-5598-2668}\,$^{1}$
	\\
	$^{1}$Shanghai Astronomical Observatory, Chinese Academy of Sciences, 80 Nandan Road, Shanghai 200030, China\\
	$^{2}$School of Astronomy and Space Science, University of Chinese Academy of Sciences, 19A Yuquan Road, Beijing 100049, China\\
	$^{3}$National Astronomical Observatories, Chinese Academy of Sciences, 20A Datun Road, Beijing 100101, China
}

\date{Accepted XXX. Received YYY; in original form ZZZ}

\pubyear{2015}

\graphicspath{{figures/}}

\usepackage{caption}
\usepackage{graphicx}
\usepackage{float} 
\usepackage{subcaption, enumerate}
\begin{document}
	\label{firstpage}
	\pagerange{\pageref{firstpage}--\pageref{lastpage}}
	\maketitle
	
	\begin{abstract}
		Detecting the cosmic 21\,cm signal from Epoch of Reionization (EoR) has always been a difficult task. Although the Galactic foreground can be regarded as a smooth power-law spectrum, due to the chromaticity of the antenna, additional structure will be introduced into the global spectrum, making the polynomial fitting algorithm perform poorly. In this paper, we introduce an improved polynomial fitting algorithm - the Vari-Zeroth-Order Polynomial (VZOP) fitting and use it to fit the simulation data. This algorithm is developed for the upcoming Low-frequency Anechoic Chamber Experiment (LACE), yet it is a general method suitable for application in any single antenna-based global 21\,cm signal experiment. VZOP defines a 24-hour averaged beam model that brings information about the antenna beam into the polynomial model. Assuming that the beam can be measured, VZOP can successfully recover the 21\,cm absorption feature, even if the beam is extremely frequency-dependent. In real observations, due to various systematics, the corrected measured beam contains residual errors that are not completely random. Assuming the errors are frequency-dependent, VZOP is capable of recovering the 21\,cm absorption feature even when the error reaches 10\%. Even in the most extreme scenario where the errors are completely random, VZOP can at least give a fitting result that is not worse than the common polynomial fitting. In conclusion, the fitting effect of VZOP depends on the structure of the error and the accuracy of the beam measurement.
	\end{abstract}
	
	
	%%%%%%%%%%%%%%%%%%%%%%%%%%%%%%%%%%%%%%%%%%%%%%%%%%
	
	%%%%%%%%%%%%%%%%% BODY OF PAPER %%%%%%%%%%%%%%%%%%
	
	\section{Introduction}
	During the final phase of the Dark Age, the first generation of stars emerged under the gravitation effect \citep{castelvecchi2019quest}. Afterwards, the first stars emit UV and X-ray photons to heat and ionize the neutral hydrogen in the universe \citep[][]{dayal2018early}. Therefore, this period is called the Cosmic Dawn (CD) and Epoch of Reionization (EoR). The cosmic 21\,cm signal which comes from the hyperfine splitting of the 1S ground state of hydrogen is an important probe to study the Dark Age, CD and EoR. Because the baryonic matter in the universe is mainly hydrogen, it is possible to detect 21\,cm signal - a forbidden line in the laboratory - by differential brightness temperature between spin temperature of hydrogen and background brightness temperature \citep{morales2010reionization, pritchard201221, zaroubi2012epoch, mesinger2019cosmic}. We always take Cosmic Microwave Background (CMB) as background to detect cosmic 21\,cm signal. When $z\lesssim200$, the difference between spin temperature and CMB temperature kept changing due to several factors such as the cosmic expansion, different ionization and recombination mechanisms, heating and cooling processes, and the evolving of the number density of different kinds of ions, showing different absorption or emission features in different periods. Therefore, the details of the pictures of the universe can be inspected through the detection of the cosmic 21\,cm signal.
	
	However, there are several difficulties in detecting 21\,cm signal. A widely accepted prediction about the biggest differential from CMB is only about 200\,mK \citep{cohen2018charting, reis2021subtlety}, while Galactic foreground and extragalactic radio sources are stronger by 4-5 orders of magnitude. In addition to the foreground, there are many other contaminants such as radio frequency interference (RFI), ground radiation temperature, and thermal noise of the receiver. All of these contaminants are always much greater than the signal.
	
	There are mainly three strategies to detect cosmic 21\,cm signal: 1) global 21\,cm signal detection, which is used by experiments including BIGHORNS \citep[Broadband Instrument for Global HydrOgen ReioNisation Signal,][]{sokolowski2015bighorns, sokolowski2015impact}, EDGES \citep[Experiment to Detect the Global EOR Signature,][]{bowman2008toward, rogers2012absolute, mozdzen2016limits, bowman2018absorption}, LEDA \citep[Large-Aperture Experiment to detect the Dark Ages,][]{greenhili2012broadband}, SCI-HI \citep[Sonda Cosmológica de las Islas para la Detección de Hidrógeno Neutro,][]{voytek2014probing}, PRI$^Z$M \citep[Probing Radio Intensity at high-Z from Marion,][]{philip2019probing}, SARAS \citep[Shaped Antenna measurement of the background RAdio Spectrum,][]{singh2017first, subrahmanyan2021saras}, CTP \citep[Cosmic Twilight Polarimeter,][]{nhan2019assessment}, REACH \citep[Radio Experiment for the Analysis of Cosmic Hydrogen][]{de2019reach}, 2) 21\,cm signal power spectrum detection, which is performed by some SKA (Square Kilometer Array) pathfinders such as 21CMA \citep[21 Centimeter Array,][]{zheng2016radio, huang2016radio}, GMRT \citep[Giant Metrewave Radio Telescope,][]{paciga2013simulation}, MWA \citep[Murchison Widefield Array,][]{beardsley2016first}, LOFAR \citep[Low-Frequency Array,][]{patil2017upper}, PAPER \citep[Precision Array for Probe the Epoch of Reionization,][]{ali201564} and HERA \citep[Hydrogen Epoch of Reionization Array,][]{deboer2017hydrogen}, 3) 21\,cm signal tomography, which is one of the main tasks of SKA in the future. However, no valuable results have been obtained by now except EDGES, which claimed an absorption profile centered at 78\,MHz with half-maximum width 19\,MHz and 0.5\,K depth \citep{bowman2018absorption}. If this profile is indeed the cosmic 21\,cm absorption feature, it will unveil new physics or astrophysics \citep[][]{chen2004spin, nelson2013moving, barkana2018possible, li2021detailed, yang2021constraints}. Nevertheless, this result is not yet widely accepted because of serious divergence from theory \citep{cohen2018charting}. SARAS studied this profile and found that it may be spectral distortion caused by EDGES low-band instrument \citep{singh2022detection}.
	
	Global 21\,cm signal detection usually uses just a single antenna to obtain the global averaged spectrum of the sky brightness temperature but ignores its spatial variation. Benefiting from the portability of single antennas, it could be relatively easy to select the experiment site. Furthermore, only a few data acquisition channels are required, and no complex data exchange network is needed. Among various Global 21\,cm signal detection experiments, only EDGES claimed that they successfully detected the signal. EDGES fitted a fifth-order polynomial to data in log space and sampled in the parameters space to detect the absorption feature \citep{bowman2018absorption}. Nth-order polynomial fitting is a common algorithm in many global cosmic 21\,cm signal detection simulations \citep[e.g.][]{pritchard2010constraining, voytek2014probing, bernardi2016bayesian, mozdzen2016limits, bowman2018absorption, gu2020direct, singh2022detection, shi2022lunar}. This method is based on the assumption that the Galactic foreground spectrum can be well described by a power-law model as it is dominated by the synchrotron radiation because of free electrons spiraling around the Galactic magnetic field lines.
	
	However, the frequency-dependent (chromatic) beam pattern introduces an artificial spectral structure, which significantly deviates from a simple power law model and hinders the reconstruction of the 21\,cm signal. Researchers typically divide the global spectrum by a beam correction factor to account for the chromaticity of the antenna. The beam correction factor is defined by
	\begin{equation}
		C=\frac{\oiint{T_b(\nu_r,\mathbf{n})B_A(\nu,\mathbf{n})\mathrm{d}\Omega}}{\oiint{T_b(\nu_r,\mathbf{n})B_A(\nu_r,\mathbf{n})\mathrm{d}\Omega}},
		\label{eq:beam correction factor}
	\end{equation}
	where $T_b$ is the real sky temperature at frequency $\nu$ and direction $\mathbf{n}$, $B_A$ is the antenna beam, $\nu_r$ is the reference frequency and $\Omega$ is the solid angle \citep[e.g.][]{mozdzen2016improved, monsalve2017results, mozdzen2019spectral, sims2020testing, shen2021quantifying, anstey2022informing, spinelli2022antenna, de2022reach}. However, due to the varying spectral index of foreground temperatures in different directions, the effectiveness of the beam correction factor is often limited. Therefore, some researchers have parametrized the foregrounds \citep[][]{anstey2021general, pagano2022general}. They divided the sky maps into N regions with similar spectral indices and assumed that the spectral index within each region is constant. Assuming the antenna beam can be precisely measured, they employed a Bayesian approach to estimate the spectral indices of each region, allowing for a more accurate foreground subtraction. This result can be anticipated since it incorporates the information of the antenna beam into the signal reconstruction model. 
	
	Similarly, the algorithm described in this paper leverages the same principle and further takes into account the scenario of imprecise beam measurements. We develop the Nth-order polynomial fitting method above, i.e. Vari-Zeroth-Order Polynomial (VZOP hereafter) fitting, which will be used in a ground-based global 21\,cm signal experiment named "Low-frequency Anechoic Chamber Experiment" \citep[LACE][]{LACE} in the future. LACE plans to build a low-frequency underground anechoic chamber to provide a radio-quiet environment, the schematic of the chamber is shown in the Fig.\,1 of \citet{LACE}. We intend to apply the VZOP in LACE in the future. Consequently, some of the parameters chosen for our simulation are aligned with the context of LACE. Nevertheless, it is crucial to note that VZOP is not exclusively tailored for LACE; rather, it is a general algorithm and can be employed effectively in any single antenna-based global 21\,cm signal experiments.

	The rest of this paper is organized as follows. Section~\ref{sec:MaM} presents the foreground, cosmic 21\,cm absorption feature and noise model used in the simulation, and details the VZOP algorithm. In Section~\ref{sec:Simulation} we describe the simulation process. We provide the fitting results assuming that the beam is accurately known in Section~\ref{sec:FT} and that the beam has errors in Section~\ref{sec:IABP}. We discuss and conclude in section~\ref{sec:Discussions and Conclusions}.

	\section{Models and Methods}
	\label{sec:MaM}
		
	\subsection{Foreground model}	
	In this work, we make use of the GSM \citep[Galactic Global Sky Model,][]{de2008model}, which provides a sky maps generator that can generate maps at any frequency between 10\,MHz and 10\,GHz. GSM used the first 3 principal components out of 11 components obtained using PCA in the sky region with observed data at all 11 frequencies, then fitted 3 principal components with a cubic spline at any frequency to get all-sky maps. It should be noted that the foreground emission is slightly polarized, which is direction- and frequency-dependent \citep{spinelli2019contamination}. This could also create an artificial spectral structure in the measured global spectra. In this work, we do not consider this factor to keep the conciseness. Further investigation into characterising this factor will be the aim of future work.
	
	\subsection{Cosmic 21\texorpdfstring{$\,$}{}cm signal models}
	This work focuses on the absorption profile in $50-120$\,MHz, which has the largest amplitude and is the most significant signature of the global EoR 21\,cm signal. In this paper, we choose two signal models as input to simulate the expected 21\,cm absorption feature: the \textbf{Gaussian model} refers to a simple Gaussian profile \citep{LACE}, and the \textbf{EDGES model} refers to a flattened Gaussian profile found by EDGES \citep{bowman2018absorption}. The 21\,cm signal brightness temperature $T_\mathrm{eor}(\nu)$ refers to the absorption or emission feature on CMB at frequency $\nu$.
	
	\textbf{Gaussian model}: This model is an approximation of the theoretical model, which can be written as
	\begin{equation}
		T_\mathrm{eor}(\nu)=A\exp\left[-\dfrac{(\nu-\nu_c)^2}{2\omega^2}\right],
		\label{eq:Gaussian model}
	\end{equation}
	where amplitude $A=-0.150$\,K, center frequency $\nu_c=78.3$\,MHz, width $\omega=5$\,MHz.
	
\begin{figure}
		\includegraphics[width=\columnwidth]{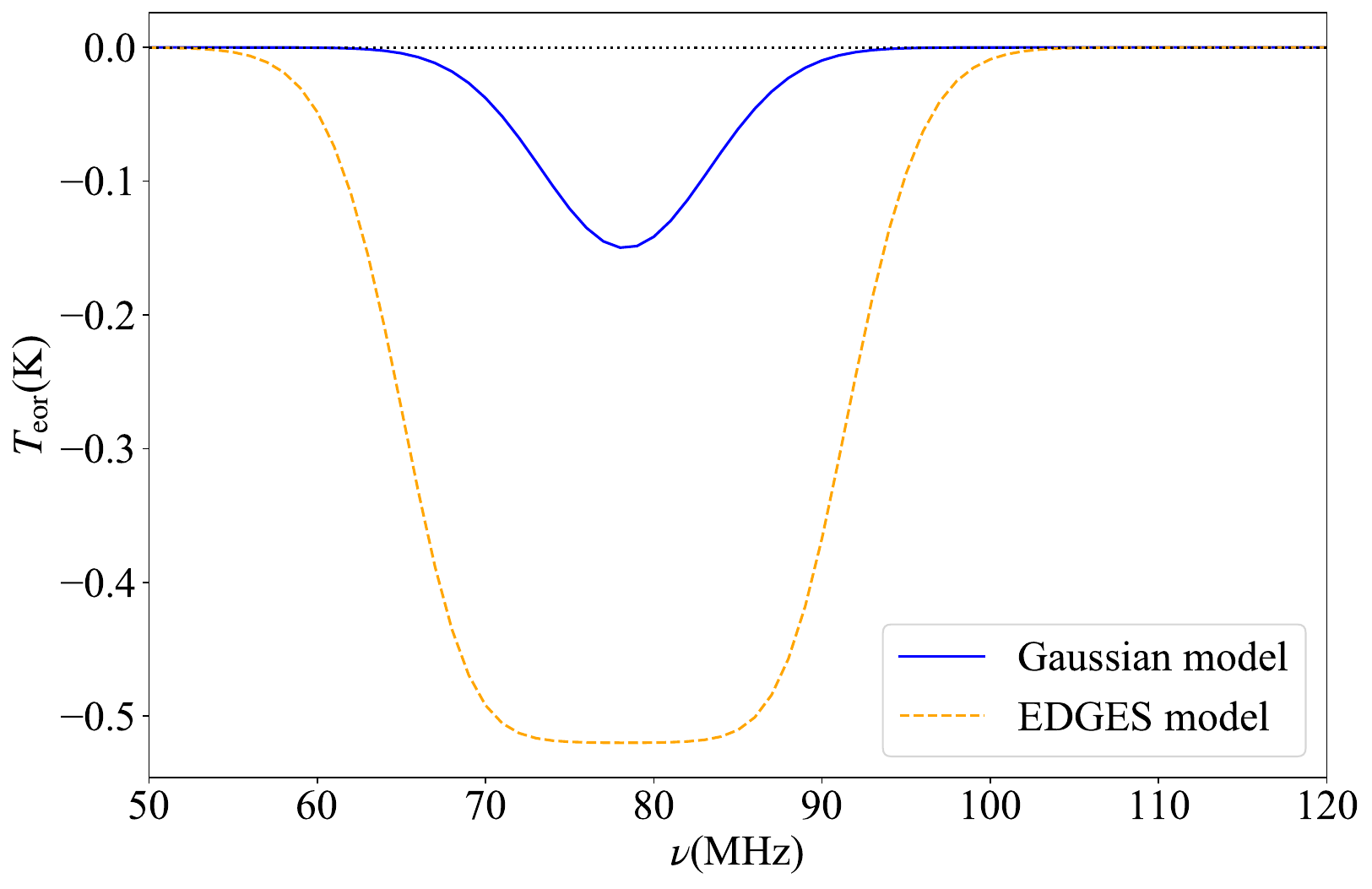}
		\caption{Brightness temperature profiles of two cosmic 21\,cm absorption feature models. The two lines represent the \textbf{Gaussian model} (blue solid) and the \textbf{EDGES model} (orange dashed), respectively.}
		\label{fig:signal model}
	\end{figure}

	\textbf{EDGES model}: EDGES have found a flattened Gaussian profile centered at $78.3$\,MHz, which can be expressed as
	\begin{equation}
		T_\mathrm{eor}(\nu)=A\left(\dfrac{1-\mathrm{e}^{-\tau\mathrm{e}^B}}{1-\mathrm{e}^{-\tau}}\right),
		\label{eq:EDGES model}
	\end{equation}
	where
	\begin{equation*}
		B=\frac{4(\nu-\nu_c)^2}{\omega^2}\log\left[-\dfrac{1}{\tau}\log\left(\dfrac{1+\mathrm{e}^{-\tau}}{2}\right)\right],
	\end{equation*}
	amplitude $A=-0.520$\,K, center frequency $\nu_c=78.3$\,MHz, full-width at half-maximum $\omega=20.7$\,MHz and flattening factor $\tau=7$. The two model profiles are shown in Fig.~\ref{fig:signal model}.

	\subsection{Thermal noise}
	We adopt a simple thermal noise model \citep{shi2022lunar} with the following standard deviation
	\begin{equation}
		\sigma_n(\nu)=\dfrac{T_A(\nu)}{\sqrt{N_\mathrm{m}\Delta\nu t_\mathrm{int}}}
		\label{eq:thermal noise},
	\end{equation}
	where $T_A$ is the antenna temperature, $N_\mathrm{m}=1$ is the number of independent measurements, and we have chosen channel bandwidth $\Delta\nu=1$\,MHz and integration time $t_\mathrm{int}=10$\,d, as such configurations ensure that the RMS remains below 10\,mK, thereby exerting no substantial influence on the signal extraction. This noise was first obtained by \citet{dicke1946measurement}. For any specific system, the enhancement in the RMS noise cannot be better than as given in equation~(\ref{eq:thermal noise}) \citep{wilson2009tools}.
	
	\subsection{Algorithm}
	\label{subsec:Algorithm}
	The first challenge to detecting the cosmic 21\,cm signal is the interference from the Galactic foreground radiation. Within 50-120\,MHz, the Galactic foreground temperature is as high as $\sim10^3$\,K, and the maximum amplitude of the absorption dip of the 21\,cm signal is just $\sim10^{-1}$\,K. Subtracting contaminants that are 4 to 5 orders of magnitude higher than the signal is extremely difficult. The antenna temperature at a certain frequency is the weighted average of the summary of the foreground and 21\,cm signals at different positions in the sky. Therefore, the antenna temperature $T_A(\nu)$ at $\nu$ contains a frequency-dependent antenna beam, which can be expressed as
	\begin{equation}
		T_A(\nu)=\frac{\oiint{T_b(\nu,\mathbf{n})B_A(\nu,\mathbf{n})\mathrm{d}\Omega}}{\oiint{B_A(\nu,\mathbf{n})\mathrm{d}\Omega}}\neq \dfrac{1}{4\pi}\oiint{T_b(\nu,\mathbf{n})\mathrm{d}\Omega}=\bar{T}_b(\nu).
		\label{eq:antanna temperature}
	\end{equation}
	where $\bar{T}_b(\nu)$ is the arithmetic mean of the observed sky temperature (i.e. assuming a uniform isotropic antenna).
	
	The common algorithm is to do a polynomial fit in logarithmic space.
	\begin{equation}
		\hat{T}_A(\nu)=\hat{T}_A(\nu_r)\exp{\left[\sum_{n=1}^Na_n\log^n(\dfrac{\nu}{\nu_r})\right]},
		\label{eq:polynomial fitting}
	\end{equation}
	where the hat indicates a modelled quantity, zeroth-order term $\hat{T}_A(\nu_r)$ is the modelled value of the antenna temperature at reference frequency $\nu_r$, and $a_n$ is the coefficient of the $n$th-order term. If the antenna temperature spectrum does contain the EoR component, then the residuals from the antenna temperatures minus the fitted values will show some structures. Next, assuming a 21\,cm absorption feature model, adding a polynomial model and refitting, we have
	\begin{equation}
		\hat{T}_A(\nu)=\hat{T}_A(\nu_r)\exp{\left[\sum_{n=1}^Na_n\log^n(\dfrac{\nu}{\nu_r})\right]}+\hat{T}_\mathrm{eor}(\nu).
		\label{eq:polynomial fitting with signal}
	\end{equation}
	If the residuals obtained by subtracting the fitted values from the antenna temperatures become small and uniform, the fitted signal parameters are considered to be true parameters. However, this algorithm usually does not perform well. The chromaticity of the antenna introduces additional structure in the temperature spectrum, which makes the reconstructed signal deviate from the expected signal \citep[e.g.][]{wilensky2022exploring, mozdzen2016limits}. This requires researchers to carefully design the antenna and handle the effects of chromaticity \citep[e.g.][]{shi2022lunar, de2008model, subrahmanyan2021saras, philip2019probing, scheutwinkel2022bayesian, shen2022bayesian}. The problem introduced by the chromatic antenna beam could be partly solved by containing the frequency dependence of the beam pattern in the solution process. 
	
	In this paper, we use VZOP to improve signal reconstruction by fitting chromaticity-generated structures to an improved polynomial model. Averaging the beam and sky temperature over 24 hours to define a 24-hour averaged beam model $\bar{B}(\nu,\theta)$ and a 24-hour averaged temperature model $\bar{T_b}(\nu,\theta)$, where $\theta$ is the declination. Then, we can obtain the global averaged temperature as follows: 
	\begin{equation}
		\begin{split}
			\bar{T}_A(\nu)
			&=\dfrac{1}{2\pi}\iiint{B(\nu,\theta,\phi)T_b(\nu,\theta,\phi-\phi')\cos{\theta}\mathrm{d}\theta\mathrm{d}\phi\mathrm{d}\phi'} \\
			&=\int{\bar{T}_b(\nu,\theta)\bar{B}(\nu,\theta)\cos{\theta}\mathrm{d}\theta},
			\label{eq:global averaged temperature}
		\end{split}
	\end{equation}
	where $\phi$ is the right ascension, the directional antenna gains below the horizon are set to be 0. $\bar{T_b}(\nu,\theta)$ and $\bar{B}(\nu,\theta)$ are just dependent on frequency $\nu$ and $\theta$ because the integral relating to $\phi$ is eliminated after 24 hours averaging. The Galactic foreground is regarded to be roughly a power law spectrum, and the antenna temperature without the signal can be approximated as a separable variable, i.e. $T(\nu,\mathbf{n})=T(\nu/\nu_r)\times T(\mathbf{n},\nu_r)$. Therefore, the modeled value of $\bar{T_b}(\nu,\theta)$ can be written as a polynomial in log space
	\begin{equation}
		\begin{split}
			\hat{\bar{T}}_b(\nu,\theta)
			&=\hat{\bar{T}}_\mathrm{gal}(\nu_r,\theta)\exp{\left[\sum_{n=1}^Na_n\log^n(\dfrac{\nu}{\nu_r})\right]}+\hat{\bar{T}}_\mathrm{eor}(\nu) \\
			&=\hat{\bar{T}}_\mathrm{gal}(\nu,\theta)+\hat{\bar{T}}_\mathrm{eor}(\nu)
			\label{eq:sky temperature},
		\end{split}
	\end{equation}
	where $\hat{\bar{T}}_\mathrm{gal}(\nu,\theta)$ is the modelled value of mean Galactic brightness temperature averaging over 24 hours, $\hat{\bar{T}}_\mathrm{eor}(\nu)$ is the modelled value of 21\,cm signal mean brightness temperature averaging over 24 hours (21\,cm signal is independent of $\theta$), and we set $\nu_r=85$\,MHz. We define
	\begin{equation}
		S(\nu;\nu_r,\mathbf{a})\equiv \exp\left[\sum_{n=1}^Na_n\log^n(\dfrac{\nu}{\nu_r})\right]
		\label{eq:polynomial}.
	\end{equation}
	According to equation~(\ref{eq:global averaged temperature}), (\ref{eq:sky temperature}) and (\ref{eq:polynomial}), discretizing angle and frequency
	\begin{equation}
		\begin{split}
			\hat{\bar{T}}_A(\nu_i)
			&=\int{\hat{\bar{T}}_b(\nu_i,\theta)\bar{B}(\nu_i,\theta)\cos{\theta}\mathrm{d}\theta} \\
			&\approx\sum_j\hat{\bar{T}\textsc{}}_\mathrm{gal}(\nu_r,\theta_j)S(\nu_i;\nu_r,\mathbf{a})\bar{B}(\nu_i,\theta_j)\cos\theta_j+\hat{\bar{T}}_\mathrm{eor}(\nu_i).
			\label{eq:discretization}
		\end{split}
	\end{equation}
	Writing equation~(\ref{eq:discretization}) as matrix form
	\begin{equation}
		\hat{\mathbf{T}}_A=\mathbf{S}(\mathbf{a})\mathbf{B}\hat{\mathbf{T}}_\mathrm{gal}+\hat{\mathbf{T}}_\mathrm{eor},
		\label{eq:matrix}
	\end{equation}
	where
	\begin{equation*}
		\begin{cases}
			B_{i,j}\equiv\bar{B}(\nu_i,\theta_j)\cos{\theta_j}, \\
			\mathbf{S}(\mathbf{a})=\mathrm{diag}[S(\nu_1;\nu_r,\mathbf{a}),S(\nu_2;\nu_r,\mathbf{a})\ldots], \\
			\mathbf{T}_A=[T_A(\nu_1), T_A(\nu_2)\ldots]^T, \\
			\mathbf{T}_\mathrm{gal}=[T_\mathrm{gal}(\nu_r,\theta_1), T_\mathrm{gal}(\nu_r,\theta_2)\ldots]^T, \\
			\mathbf{T}_\mathrm{eor}=[T_\mathrm{eor}(\nu_1), T_\mathrm{eor}(\nu_2)\ldots]^T. \\
		\end{cases}
	\end{equation*}
	The $T_\mathrm{gal}$ is a vector containing real Galactic temperatures at different declinations at frequency $\nu_r$. 
	
	In this paper, we take $N_\mathrm{bin}$ bins at equal intervals between the declination interval [$\theta_h$, $\theta_l$] (If the observation site is in the northern hemisphere, $\theta_h=90^\circ$, $\theta_l$ is the southernmost declination that can be seen; if it is in the southern hemisphere, $\theta_h$ is the northernmost declination that can be seen, $\theta_l=-90^\circ$.). In this paper, we use a 5th-order VZOP to fit our mock data, and unless otherwise specified, $N_\mathrm{bin}=10$.
	
	We use a truncated likelihood as our posterior distribution (The upper and lower bounds for each parameter are shown in Table~\ref{tab:prior})
	\begin{multline}
		\log p(\mathbf{a},\hat{\mathbf{T}}_\mathrm{gal},\mathbf{p}_\mathrm{eor}\vert\mathbf{T}_A)= -\dfrac{1}{2}\left[N_{\nu}\log(2\pi)-\log(\mathrm{det}\mathbf{\Sigma})\right] \\
		-\dfrac{1}{2}\left[\left(\mathbf{T}_A-\mathbf{S}(\mathbf{a})\mathbf{B}\hat{\mathbf{T}}_\mathrm{gal}-\hat{\mathbf{T}}_\mathrm{eor}\right)^T\mathbf{\Sigma}^{-1}\left(\mathbf{T}_A-\mathbf{S}(\mathbf{a})\mathbf{B}\hat{\mathbf{T}}_\mathrm{gal}-\hat{\mathbf{T}}_\mathrm{eor}\right)\right],
		\label{eq:likelihood}
	\end{multline}
	where $N_\nu=71$ is the number of frequency channels, and $\mathbf{\Sigma}$ is a diagonal matrix because we assume that the thermal noise of different frequency channels is independent, and the $i$-th element on the diagonal is
	\begin{equation*}
		\Sigma_{i,i}=\dfrac{T_A(\nu_i)}{\sqrt{t_\mathrm{int}\Delta\nu}}.
	\end{equation*}
	Even if the assumed frequency independence does not hold, VZOP can still be used because it does not rely on this assumption. In addition, the first term on the right side of equation~(\ref{eq:likelihood}) is the normalization coefficient. Although this term can be ignored during Bayesian parameter estimation, it is necessary to include it when computing the Bayesian evidence afterwards.
	\begin{table}
		\centering
		\caption{The upper and lower bounds for each parameter in the uniform prior distribution.}
		\label{tab:prior}
		\renewcommand{\arraystretch}{1.2}
		\begin{tabular}{lccr}
			\toprule
			 & Parameters & Minimum & Maximum\\
			\hline \hline
			 & $T_1$ - $T_{N_\mathrm{bin}}$(K) & $0$ & $10000$ \\
			 & $a_1$ & $-3$ & $-2$ \\
			 & $a_2$ - $a_5$ & $-2.5$ & $2.5$ \\
			\hline
			\multirow{3}{*}{\textbf{Gaussian model}} & $A$(K) & $-1$ & $0$\\
			& $\nu_c$(MHz) & $50$ & $120$\\
			& $\omega$(MHz) & $0$ & $70$\\
			\hline
			\multirow{4}{*}{\textbf{EDGES model}} & $A$(K) & $-1$ & $0$\\
			& $\nu_c$(MHz) & $50$ & $120$\\
			& $\omega$(MHz) & $0$ & $200$\\
			& $\tau$ & $-70$ & $70$\\
			\bottomrule
		\end{tabular}
	\end{table}

	Theoretically, VZOP would outperform polynomial fitting algorithms. The principle of VZOP is that since the chromaticity of the antenna cannot be avoided, the distribution of the foreground temperature is simulated by sampling 24-hour averaged temperatures in several declination bins. From a mathematical point of view, VZOP is equivalent to an improved polynomial fitting model with the zeroth-order term that varies with frequency, and the simplified expression is (according to equation~(\ref{eq:sky temperature}))
	\begin{equation*}
	    \hat{\bar{T}}_A(\nu_i) = \sum_{j=1}^{N_\mathrm{bin}}\hat{\bar{T}}_b(\nu_i,\theta_j)=\exp{\left[a_0(\nu_i)+\sum_{n=1}^Na_n\log^n(\dfrac{\nu_i}{\nu_r})\right]}+\hat{\bar{T}}_\mathrm{eor}(\nu_i).
	\end{equation*}
	The degree of freedom of the zeroth-order terms is equal to the rank of matrix $\mathbf{B}$. When the number of frequency channels is fixed (i.e. the number of rows of B is fixed), if the rank of $\mathbf{B}$ increases with declination bins (i.e. the number of columns of $\mathbf{B}$ increases), the fitting result will be better theoretically. The maximum number of declination bins should not exceed the number of frequency channels. For the convenience of discussion, all matrices $\mathbf{B}$ constructed in this paper are full-rank matrices, and the number of columns is not greater than the number of rows (i.e. the number of declination bins is not greater than frequency channels). Formally, VZOP can also obtain the 24-hour averaged temperature within each declination bin, but we are not interested in this nuisance parameter.
	
	We use a python package \texttt{emcee} \citep{foreman2013emcee} and find the posterior is a multi-modal distribution. To find the global maximum quickly and efficiently, we use the python package \texttt{ptemcee}, which is an improved version of \texttt{emcee} using parallel tempering \citep{vousden2016dynamic}. The package \texttt{ptemcee} simultaneously samples from tempered versions of the posterior distribution $p(\theta)$,
	\begin{equation*}
		p_T(\theta)\propto L(\theta)^{1/T}\pi(\theta),
	\end{equation*}
	where $L(\theta)$ and $\pi(\theta)$ are respectively likelihood and prior distributions. The probability distribution at the lowest temperature $T=1$ corresponds to the target posterior distribution, and as the temperature increases, the distribution gradually approaches the prior distribution. At the highest temperature, \texttt{ptemcee} approximates sampling from the prior distribution. In this way, the Markov chain at low temperatures is responsible for exploring a specific mode, while the Markov chain at high temperatures can explore the entire posterior space. By exchanging information between different temperatures, \texttt{ptemcee} can effectively sample the high-dimensional and multi-modal distribution.
	
	\section{Simulation}
	\label{sec:Simulation}
	In our simulation, we positioned the antenna at the site of the 21CMA ($42.93^\circ$N, $86.68^\circ$E), which is far away from artificial radio sources.
	
\begin{figure*}
		\centering
		\includegraphics[width=2\columnwidth]{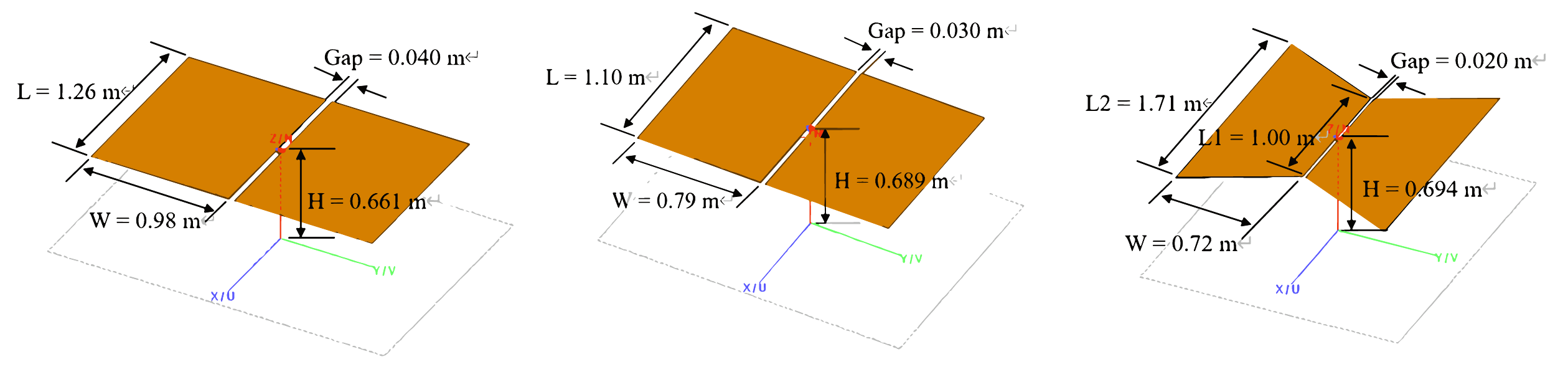}
		\caption{The simulated images of nominal blade antenna (left), optimized blade antenna (middle) and bow-tie blade antenna (right).}
		\label{fig:antennas}
	\end{figure*}

\begin{figure*}
		\centering
		\includegraphics[width=\linewidth]{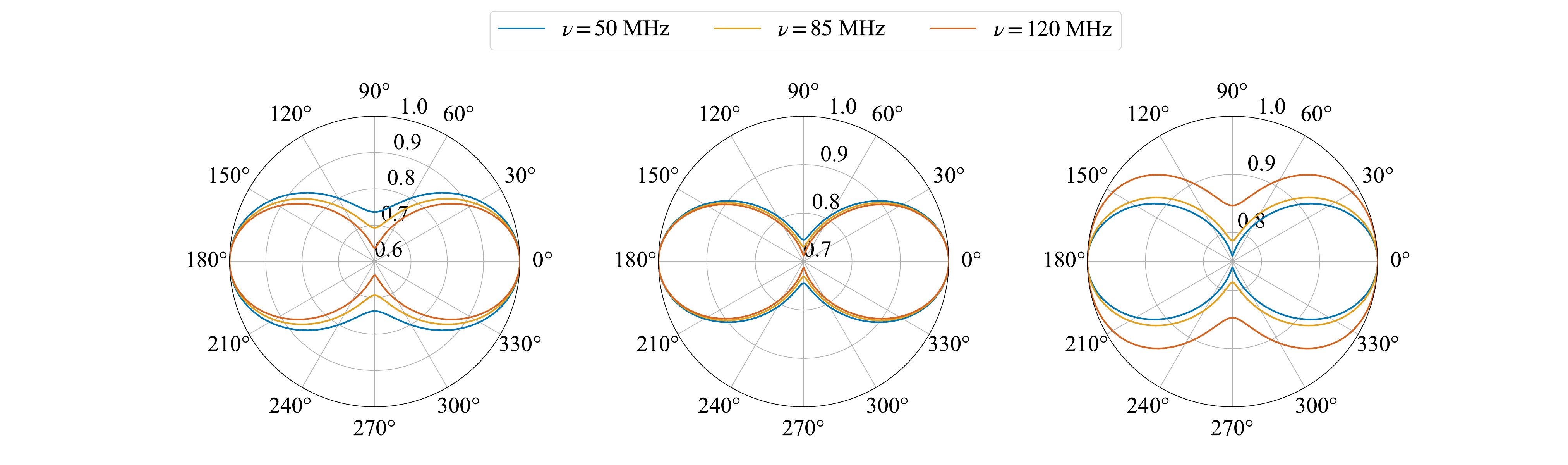}
		\caption{The cross section of the power beam profiles at $\Theta=30^\circ$ at 50\,MHz (blue), 85\,MHz (orange) and 120\,MHz (vermilion). The three panels are the nominal blade antenna (left), optimized blade antenna (middle) and bow-tie blade antenna (right), respectively.}
		\label{fig:chromaticity}
	\end{figure*}

	The antenna to be used in LACE has not been finalized at this time. An important principle in antenna design is to minimize the antenna chromaticity and to avoid the coupling of spatial structures into the frequency domain as much as possible. Three candidate antenna designs are to be tested: a nominal blade antenna, an optimized blade antenna and a bow-tie blade antenna. Each of them consists of two identical Perfect Electric Conductor (PEC) panels with a thickness to be 2\,mm. They work in frequency bands $50-120$\,MHz corresponding to redshifts at $z=10-27$ where the strong absorption feature is expected to exist. The ground plane is assumed to be an infinite uniform PEC. The structures of the three simulated antennas are shown in Fig.~\ref{fig:antennas}. The cross-section of antenna beam profiles, cut at zenith angle $\Theta=30^\circ$ with different azimuth angle $\Phi$, are shown in Fig.~\ref{fig:chromaticity}.
	
	Under the assumption that the antenna beams do not change with time, we use the following method to generate mock observation data i.e. the antenna temperature.
	\begin{enumerate}[1)]
		\item Given an arbitrary frequency, the GSM model can generate a global map using \texttt{healpix} RING scheme with an angular resolution of approximately $6.87^\prime$ ($n_\mathrm{side}=512$) \citep{HEALPixDescription}.
		\item We divide the sky map into 1024 bins along the declination of the celestial coordinate system and take the averaged temperature in each bin as the 24-hour averaged brightness temperature at the declination of the centre of the bin.
		\item We could perform the same for the antenna beam, but a preprocessing is required because the simulated beam is not equal in area but equal in angle. The zenith and azimuth angles of our generated antenna gains are both $2^\circ$ apart (the gains below the horizon are set to 0), so the resolution is much lower than $6.87^\prime$. Firstly, therefore, we calculate the smallest $n_\mathrm{side}=16$ at which there is at least one gain within each pixel. We obtain the arithmetic mean if there is more than one gain within each pixel. Secondly, we perform bilinear interpolation based on the nearest four points and expand it to a map with $n_\mathrm{side}=512$. 
		\item The antenna is placed at the observation site at a certain moment (the time can be chosen arbitrarily because it does not affect the final result). Next, similar to step 2), the gains within each bin are summed and averaged to obtain the 24-hour averaged beam.
		\item Different from the common polynomial fitting, if we want to generate the antenna temperature by multiplying the 24-hour averaged temperatures and the 24-hour averaged beam, we need to multiply the beam by the cosine of the corresponding declination and then normalize, and then the result of the multiplication is the antenna temperature as following,
		\begin{equation*}
			\bar{T}_A(\nu_i)=\frac{\sum_{j=1}^{1024}\bar{T}\textsc{}_\mathrm{gal}(\nu_i,\theta_j)\bar{B}(\nu_i,\theta_j)\cos\theta_j}{\sum_{j=1}^{1024}\bar{B}(\nu_i,\theta_j)\cos\theta_j}.
	    \end{equation*}
	\end{enumerate}
	
\begin{figure}
	    \includegraphics[width=\columnwidth]{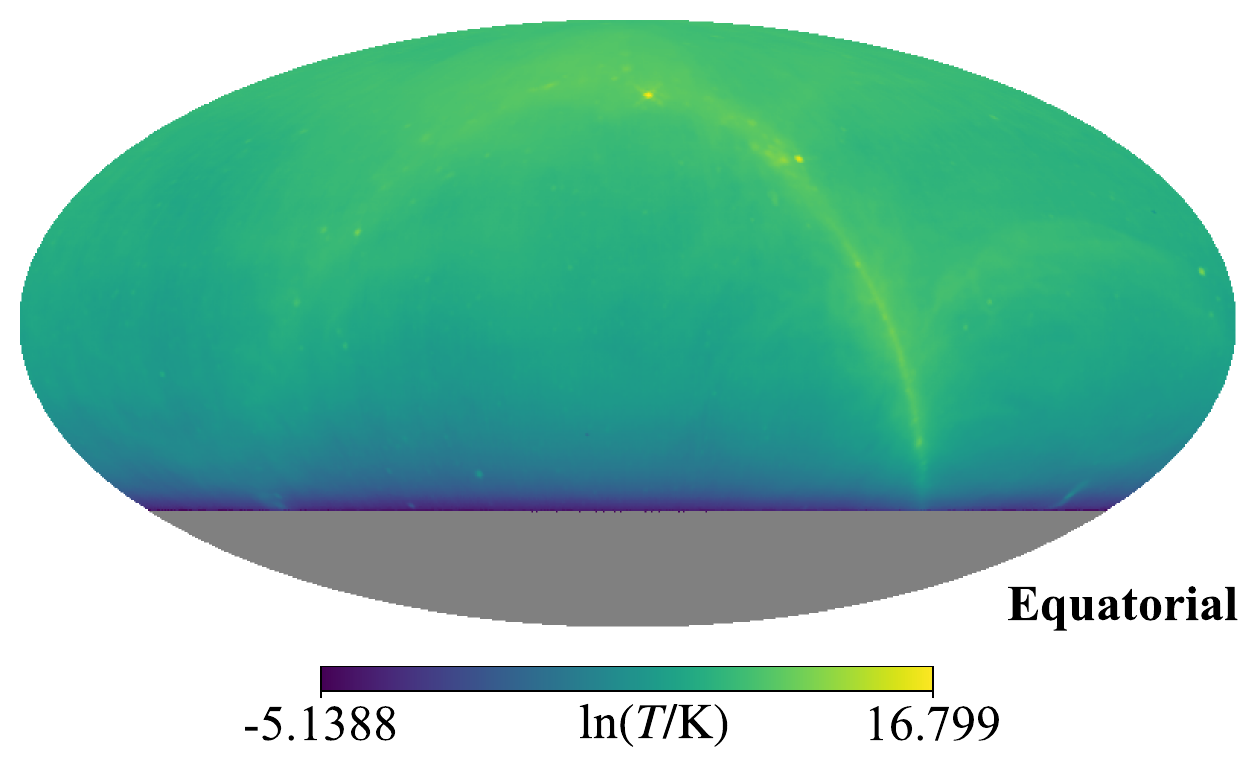}
		\caption{Sky map weighted by 24-hour averaged beam (the maximum is normalized to 1) of the optimized blade antenna at 85\,MHz.}
		\label{fig:three_antennas_maps}
	\end{figure}

	In Fig.~\ref{fig:three_antennas_maps}, we show the sky brightness temperature weighted by the 24-hour averaged antenna beam of the optimized blade antenna at 85 MHz. The temperature distribution of the figure is independent of the longitude of the observation site. The figure shows the map using our defined 24-hour averaged antenna beam after taking into account the effects of Earth's shading. It tells us which part of the sky has made the main contribution to observational data. The darker area represents the area of the sky near the southern horizon that is only visible for part of the day. The grey area at the bottom represents the area below the horizon throughout the day.
	
	\section{Fitting Results}
	\label{sec:FT}
	In this section, we performed signal reconstruction at the 21CMA site assuming that the antenna beam can be precisely measured. 
	
	\subsection{Common Polynomial Fitting}
\begin{figure}
		\includegraphics[width=\columnwidth]{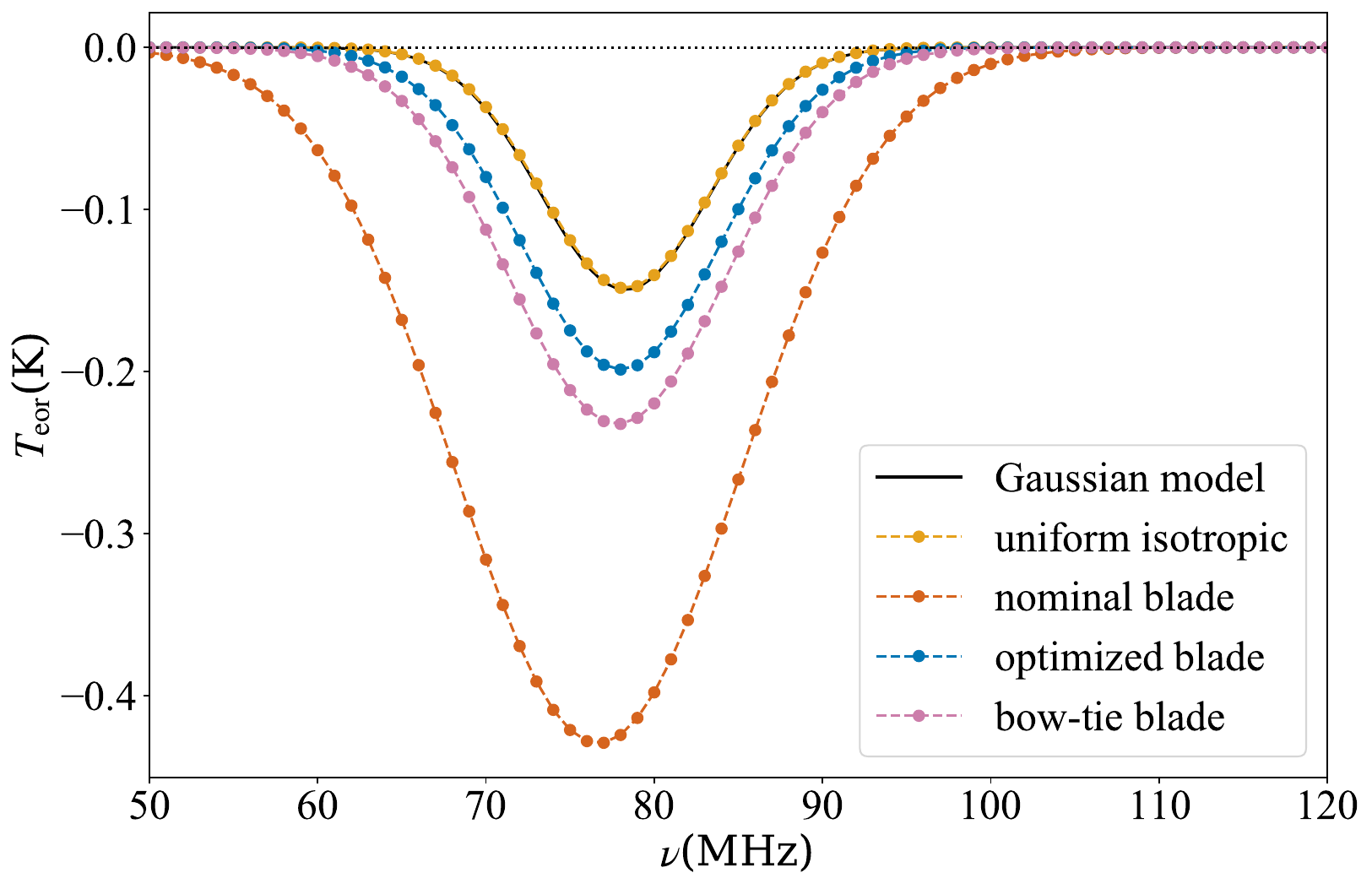}
		\caption{Fitting results of polynomial fitting for \textbf{Gaussian model}. The black solid line represents the input \textbf{Gaussian model}. Four dashed lines with marker represent the results based on the uniform antenna (orange), nominal blade antenna (vermilion), optimized blade antenna (blue) and bow-tie blade antenna (reddish purple), respectively.}
		\label{fig:polynomial fitting}
	\end{figure}

	Before using the VZOP fitting, we first present the result of common polynomial fitting. Fig~\ref{fig:polynomial fitting} shows the results obtained by fitting the antenna temperatures generated by different antennas using the polynomial fitting (\textbf{Gaussian model}), where a uniform isotropic antenna is added for comparison. When using the uniform isotropic antenna, it is not surprising that the reconstructed absorption feature agrees well with the input feature. When using other antennas, the results of signal reconstruction are relatively poor due to the existence of chromaticity.

	Since the uniform isotropic antenna is achromatic, it is predictable that the reconstructed absorption feature and the input feature are perfectly coincident In Fig~\ref{fig:polynomial fitting}. However, the fitting results based on the mock data obtained from the other three antennas have different degrees of deviation. The optimized blade antenna has the least deviation, while the nominal blade antenna has the strongest deviation from the input absorption feature. This is consistent with Fig~\ref{fig:chromaticity}, where the frequency dependence of the optimized blade antenna beam is the weakest, and that of the nominal blade antenna is the strongest. 
	
	It can be seen from the above results that the use of non-ideal antennas will lead to poorer performance of polynomial fitting to extract the 21\,cm absorption feature. The stronger the frequency dependence of the antenna, the worse the performance of the polynomial fitting. Because of the different foreground temperatures at different points in the sky, the spatial variation of the beam with frequency will couple into the spectrum, which will eventually lead to additional structure in the spectrum and affect the final fitting result.
	
	\subsection{Mitigating the Chromatic Effect with VZOP}
	\label{subsec:Effects of Chromaticity on VZOP}
\begin{figure}
		\includegraphics[width=\columnwidth]{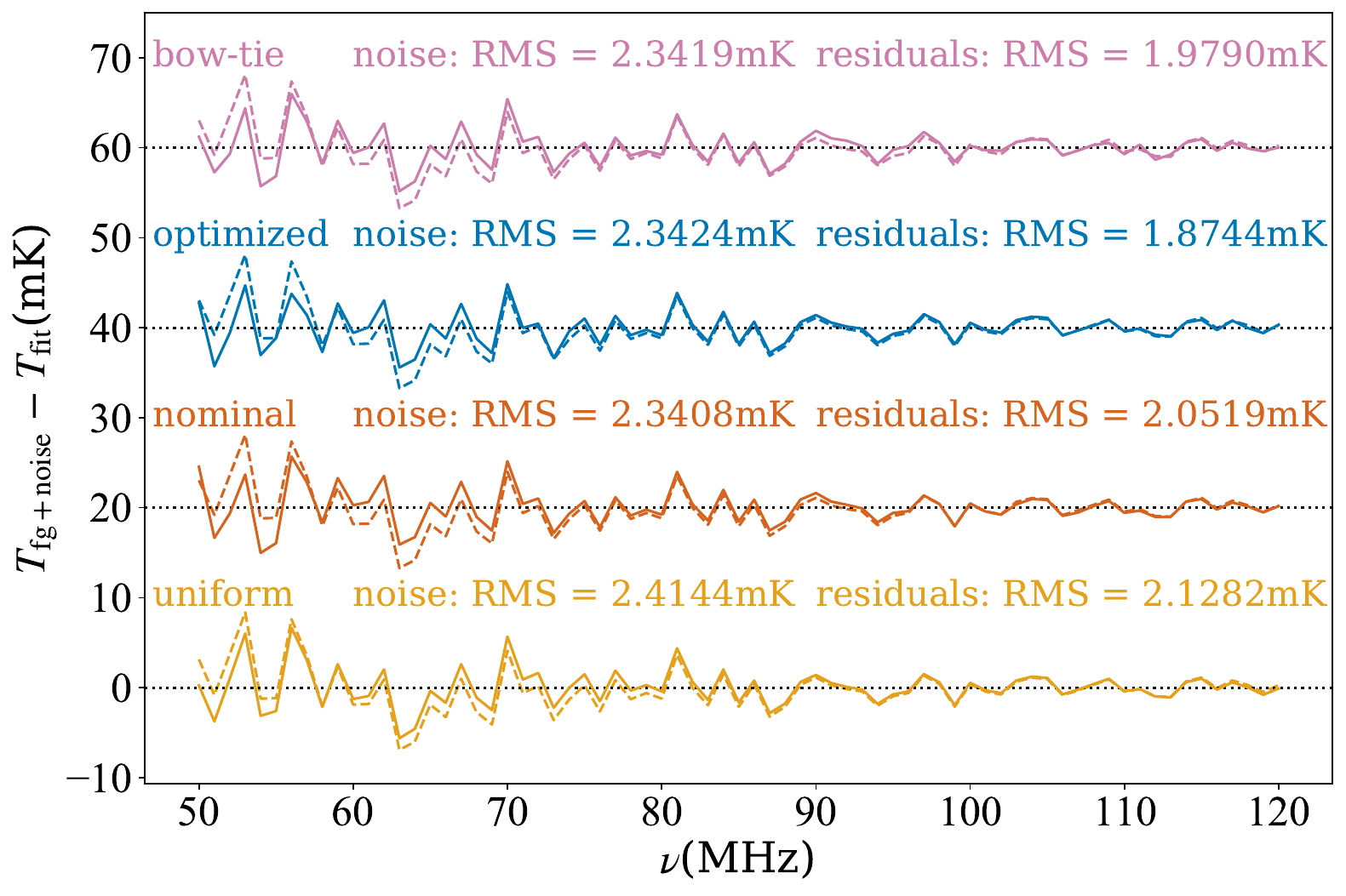}
		\caption{The residuals after fitting foreground with only noise for different antennas. The dashed lines represent the input noise and the solid lines are residuals. From the bottom up, the lines with different colours represent residuals for uniform isotropy (orange), nominal blade (vermilion), optimized blade (blue) and bow-tie blade antenna (reddish purple), respectively. The lines are shifted vertically for clarity.}
		\label{fig:fit structure}
	\end{figure}

	The chromaticity will introduce additional structure on the 24-hour averaged antenna temperature spectrum and have an impact on signal reconstruction. Therefore, we first fit the foreground model with only noise (i.e. no 21\,cm signal) using VZOP. The fitting results are shown in Fig~\ref{fig:fit structure}. For comparison, we also introduce a uniform isotropic antenna. The lines are shifted vertically for clarity. Since we generate the noise using a random number generator with the same seed for all antennas, the shapes of these lines are very similar. It can be seen that the residuals after fitting the foreground are similar to that of input noise. The Root Mean Squares (RMSs) of the residuals of these lines are less than 2\,mK, far less than $\sim10^{2}$\,mK of the 21\,cm signal.

	The RMS of a uniform antenna is small and its residuals are uniform, which is reasonable because the uniform antenna has no chromaticity. The RMSs of the other three antennas are similar to that of the uniform antenna, indicating that VZOP effectively fits the structure introduced by chromaticity into the improved polynomial model.
	
	\subsection{VZOP Fitting}
	\label{subsec:VZOP_fitting}
\begin{figure*}
		\centering 
		\includegraphics[width=\linewidth]{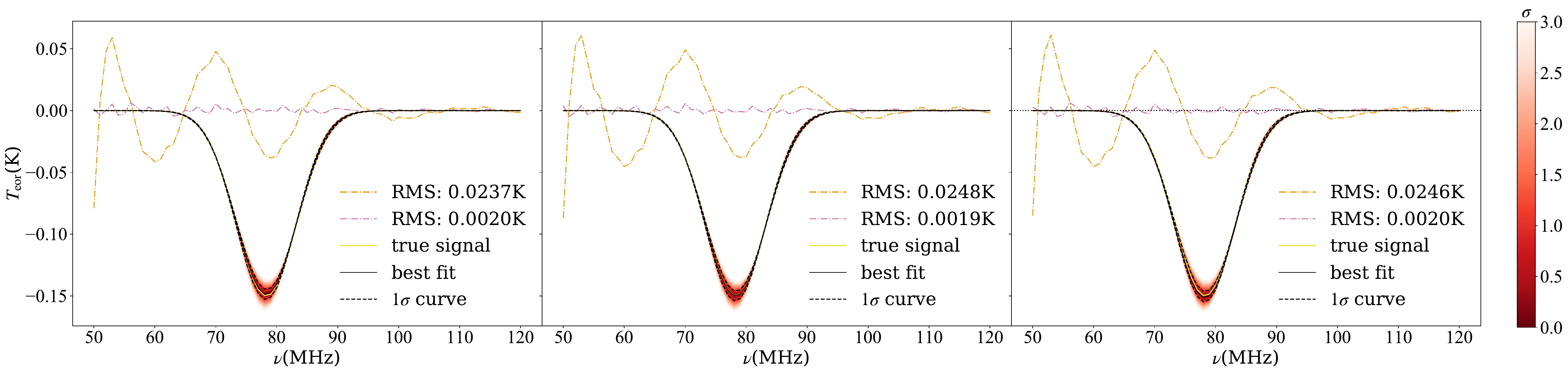}
		\caption{Fitting results for \textbf{Gaussian model} with nominal blade (left), optimized blade (middle) and bow-tie blade antenna (right). The orange dash-dotted line denotes the residuals after fitting and removing only the foreground, while the reddish-purple dash-dotted line represents the residuals after fitting and removing both foreground and absorption features. The yellow solid line denotes the input absorption feature, while the black solid is the recovered best-fit line. Also, we plot dozens of lines to represent different error levels with different shades of colour (see the colour bar to the right of the figure). For clarity, we intentionally plot two 1$\sigma$ lines (dashed black lines).}
		\label{fig:11_recovered}
	\end{figure*}

	As described in Section~\ref{subsec:Effects of Chromaticity on VZOP}, VZOP effectively mitigates the effects of chromaticity. Therefore, in the reconstruction of the cosmic 21\,cm absorption feature, the performance of VZOP will be better than the common polynomial fit. We plot in Fig.~\ref{fig:11_recovered} the fitting results for \textbf{Gaussian model}. We can see in these three panels that if the EoR signal is included, the RMS will decrease from over 20\,mK to only about 2\,mK. The reconstructed EoR absorption features almost completely coincide with the input features for both antennas, indicating that VZOP fitting can effectively reduce the impact of the antenna chromaticity for \textbf{Gaussian model}.
	
\begin{figure*}
		\centering
		\includegraphics[width=\linewidth]{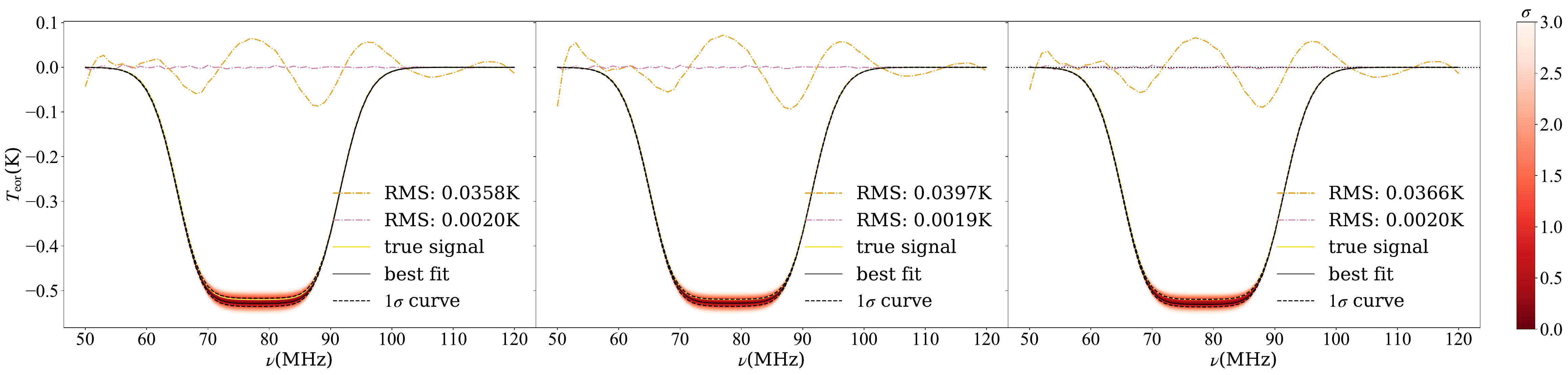}
		\caption{Fitting results for \textbf{EDGES model} with nominal blade (left), optimized blade (middle) and bow-tie blade antenna (right). The orange dash-dotted line denotes the residuals after fitting and removing only the foreground, while the reddish-purple dash-dotted line represents the residuals after fitting and removing both foreground and absorption features. The yellow solid line denotes the input absorption feature, while the black solid is the recovered best-fit line. Also, we plot dozens of lines to represent different error levels with different shades of colour (see the colour bar to the right of the figure). For clarity, we intentionally plot two 1$\sigma$ lines (dashed black lines).}
		\label{fig:22_recovered}
	\end{figure*}

	Next, we plot the fitting results for \textbf{EDGES model} in Fig.~\ref{fig:22_recovered}. We obtained similar results to Fig.~\ref{fig:11_recovered}, with the reconstructed absorption feature being highly coincident with the input feature. Compared with \textbf{Gaussian model}, the fitting result of \textbf{EDGES model} is slightly deviated, but the input absorption features of the three antennas are still all within the $1\sigma$ confidence interval of the reconstructed features. Therefore, VZOP also performs well in \textbf{EDGES model}, and the frequency dependence of the antenna does not seem to matter.

	When we use VZOP considering 10 declination bins, the input absorption feature can be recovered almost perfectly, whether it is the \textbf{Gaussian model} or the \textbf{EDGES model}. This is mainly because VZOP brings the beam's information into the polynomial model. Formally, the zeroth-order term in VZOP that varies with frequency increases the degree of freedom of the parameters. Comparing Fig.~\ref{fig:11_recovered} and Fig.~\ref{fig:22_recovered}, the best-fit result of the \textbf{Gaussian model} has less deviation from the input feature than the \textbf{EDGES model}. Because we take the set of parameters with the largest posterior probability as the best fitting value, whether it is the \textbf{Gaussian model} or the \textbf{EDGES model}, the results obtained by each run of MCMC have some fluctuations, and the fluctuation of the \textbf{EDGES model} is greater. This may be because the \textbf{EDGES model} is deeper, wider, and has more parameters so that is more difficult to constrain. But all of their best-fit values are within the 68\% confidence interval, so such fluctuations have little effect on the final results.

\begin{figure*}
		\centering
		\begin{subfigure}{2\columnwidth}
			\centering
			\includegraphics[width=\linewidth]{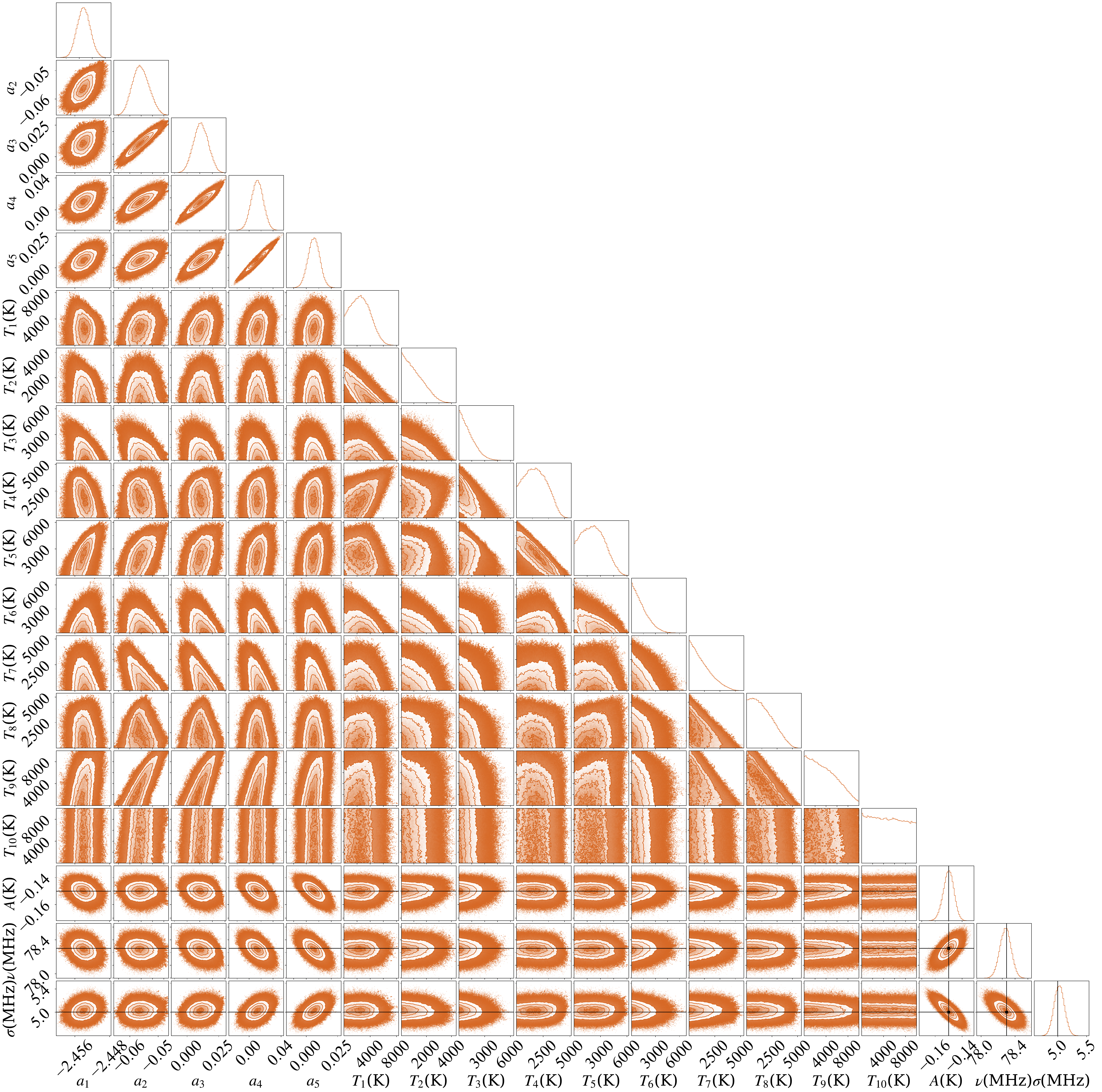}
		\end{subfigure}
		\caption{Distributions of all parameters for \textbf{Gaussian model} based on the optimized blade antenna. The parameters from left to right (from top to bottom) are 5 polynomial coefficients ($a_1-a_5$), 10 24-hour averaged temperature parameters at each bin and signal parameters ($A, \nu_c, \omega$). The true values of cosmic 21\,cm signal parameters are also shown in a black solid line.}
		\label{fig:11_opt_full_distribution}
	\end{figure*}

	The parameters distribution of the middle panel in Fig.~\ref{fig:11_recovered} (i.e. \textbf{Gaussian model} based on the optimized blade antenna) is shown in Fig.~\ref{fig:11_opt_full_distribution}. As can be seen from this figure, both the polynomial coefficients and the 21\,cm signal parameters follow normal distributions approximately. But our model has just weak constraints on the 10 24-hour averaged temperature parameters. We found that the true 24-hour averaged temperatures lie within $1\sigma$ of the best-fit value, which means we obtain the true 24-hour averaged temperatures within large error ranges. But it does not make much sense because of too large errors, we treat them as nuisance parameters and ignore them.

	We list the fitted 21\,cm signal parameters for \textbf{Gaussian model} and \textbf{EDGES model} in Table~\ref{tab:best_fit_case1} and Table~\ref{tab:best_fit_case2}, respectively. Except for $\omega$ in \textbf{EDGES model}, all input parameters are within the $1\sigma$ error interval of the recovered parameters. To verify the plausibility of the model themselves, we also calculated the Bayesian evidence using \texttt{polychord} \citep{polychord2015a, polychord2015b}, which are also placed in two tables. It can be seen from these two tables that if the 21\,cm absorption feature is considered during sampling, the evidence significantly increases while the RMSs significantly decrease. This indicates that the signal reconstruction results of VZOP are highly reliable.
	\begin{table*}
		\centering
		\caption{The best-fit parameters with 1$\sigma$ error, values of RMS and Bayesian evidence for \textbf{Gaussian model}. $\mathrm{log}Z$ is the logarithm of the evidence and RMS represents the root-mean-square. The subscript "fore" denotes the result fitting only the foreground.}
		\label{tab:best_fit_case1}
		\renewcommand{\arraystretch}{1.2}
		\begin{tabular}{lccccccccr}
			\toprule
			\textbf{Gaussian model} & Antenna & $\mathrm{log}Z_\mathrm{fore}$ & $\mathrm{log}Z$ & $\mathrm{RMS_{fore}(K)}$ & $\mathrm{RMS(K)}$ & $A(\mathrm{K})$ & $\nu_c(\mathrm{MHz})$ & $\omega(\mathrm{MHz})$\\
			\hline \hline
			Input & - & - & - & - & - & $-0.150$ & $78.3$ & $5.00$\\
			\hline
			\multirow{3}{*}{Recovered} & nominal blade & $-2916$ & $-463$ & $0.0238$ & $0.0020$ & $-0.1484_{-0.0040}^{+0.0039}$ & $78.28_{-0.07}^{+0.07}$ & $4.995_{-0.090}^{+0.087}$\\
			& optimized blade & $-2984$ & $-460$ & $0.0248$ & $0.0019$ & $-0.1501_{-0.0040}^{+0.0038}$ & $78.29_{-0.07}^{+0.06}$ & $5.025_{-0.088}^{+0.088}$\\
			& bow-tie blade & $-2949$ & $-461$ & $0.0246$ & $0.0020$ & $-0.1515_{-0.0034}^{+0.0051}$ & $78.29_{-0.06}^{+0.07}$ & $5.065_{-0.118}^{+0.070}$\\
			\bottomrule
		\end{tabular}
	\end{table*}
	\begin{table*}
		\centering
		\caption{The best-fit parameters with 1$\sigma$ error, values of RMS and Bayesian evidence for \textbf{EDGES model}. $\mathrm{log}Z$ is the logarithm of the evidence and RMS represents the root-mean-square. The subscript "fore" denotes the result fitting only the foreground.}
		\label{tab:best_fit_case2}
		\renewcommand{\arraystretch}{1.2}
		\begin{tabular}{lcccccccccr} % four columns, alignment for each
			\toprule
			\textbf{EDGES model} & Antenna & $\mathrm{log}Z_\mathrm{fore}$ & $\mathrm{log}Z$ & $\mathrm{RMS_{fore}(K)}$ & $\mathrm{RMS(K)}$ & $A(\mathrm{K})$ & $\nu_c(\mathrm{MHz})$ & $\omega(\mathrm{MHz})$ & $\tau$\\
			\hline \hline
			Input & - & - & - & - & - & $-0.520$ & $78.3$ & $20.7$ & $7.00$\\
			\hline
			\multirow{3}{*}{Recovered} & nominal blade & $-17209$ & $-470$ & $0.0352$ & $0.0020$ &	$-0.5272_{-0.0072}^{+0.0110}$ & $78.27_{-0.02}^{+0.03}$ & $20.78_{-0.07}^{+0.04}$ & $6.823_{-0.144}^{+0.253}$\\
			& optimized blade & $-18873$ & $-469$ & $0.0393$ & $0.0019$ & $-0.5249_{-0.0077}^{+0.0071}$ & $78.29_{-0.03}^{+0.02}$ & $20.72_{-0.04}^{+0.07}$ & $6.863_{-0.162}^{+0.180}$\\
			& bow-tie blade & $-17981$ & $-477$ & $0.0359$ & $0.0020$ & $-0.5232_{-0.0087}^{+0.0069}$ & $78.28_{-0.04}^{+0.02}$ & $20.73_{-0.04}^{+0.07}$ & $6.938_{-0.195}^{+0.165}$\\
			\bottomrule
		\end{tabular}
	\end{table*}
	
	\subsection{Chromatic Beam Effect}
\begin{figure*}
		\centering
		\includegraphics[width=\linewidth]{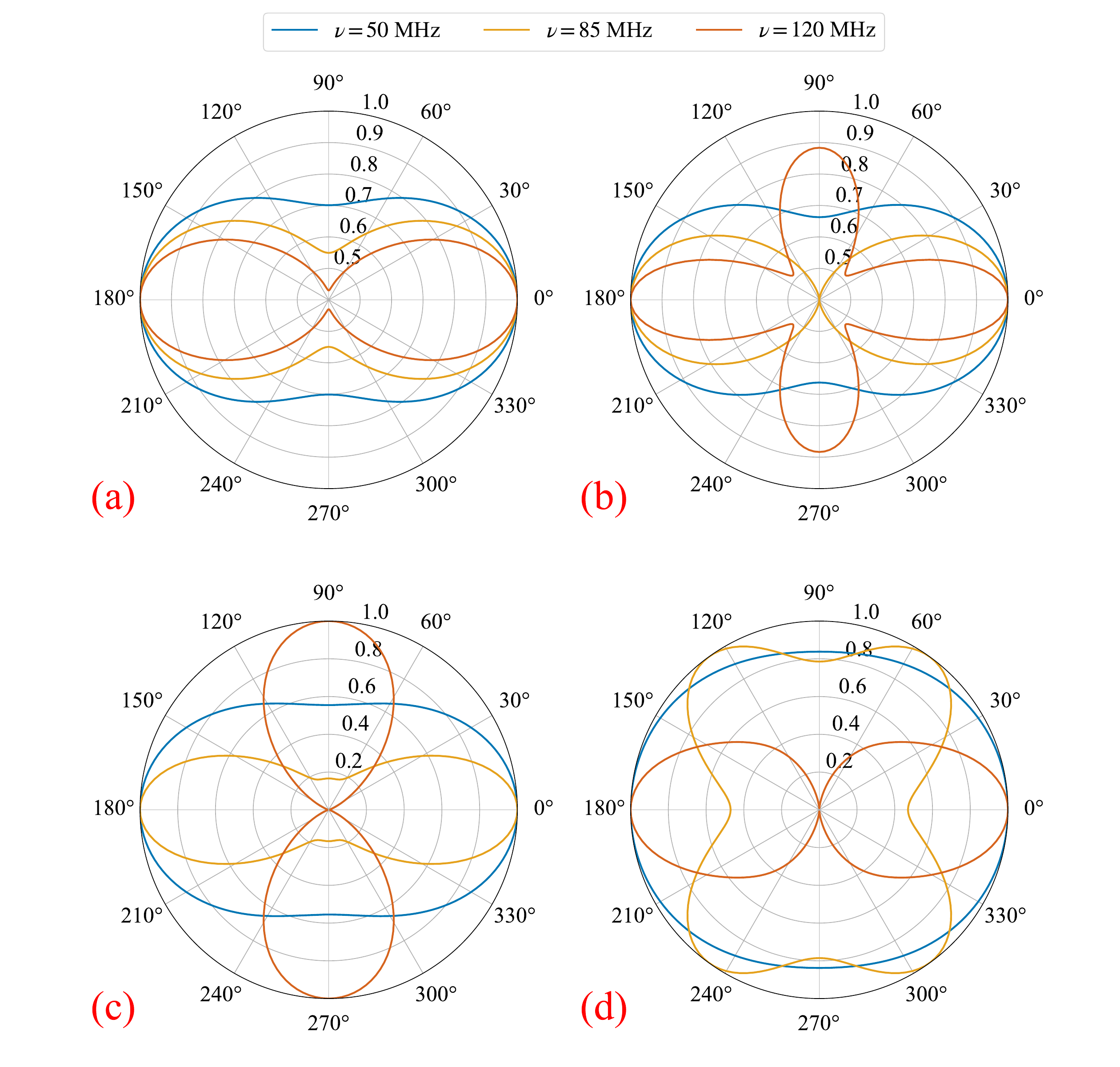}
		\caption{The cross section of the power beam profiles at $\Theta=30^\circ$ at 50\,MHz (blue), 85\,MHz (orange) and 120\, (Vermilion). Panel (a), (b) and (c) are optimized blade antennas with $W=1.25, 1.5, 2$\,m, respectively, and the panel (d) is the bow-tie blade antenna with $L=3$\,m.}
		\label{fig:extreme_chromaticity}
	\end{figure*}

	It can be seen from the above results that the VZOP fitting algorithm has an exact signal reconstruction for all three antennas and can largely reduce the impacts of chromaticity. However, the three antennas used in this paper are all close to achromatic antennas. To better understand the effect of antenna frequency dependence on VZOP, we introduce four other antennas with higher frequency-dependency, the beam pattern of which are shown in Fig~\ref{fig:extreme_chromaticity}. For the convenience of comparison, panels (a), (b) and (c) are all optimized blade antennas with different $W$ ($W=1.25, 1.5, 2$\,m, respectively), and panel (d) is the bow-tie blade antenna with $L=3$\,m. Compared with Fig~\ref{fig:chromaticity}, the four antenna beams in Fig~\ref{fig:extreme_chromaticity} vary significantly with frequency. When the $\nu=120$\,MHz, the beams in panels (b), (c), and (d) have obvious changes. The antenna in panel (b) has very obvious side lobes; the main lobe of the antenna in panel (c) is rotated $90^\circ$; the beam in panel (d) also has a great change.
	
\begin{figure*}
        \centering
		\includegraphics[width=\linewidth]{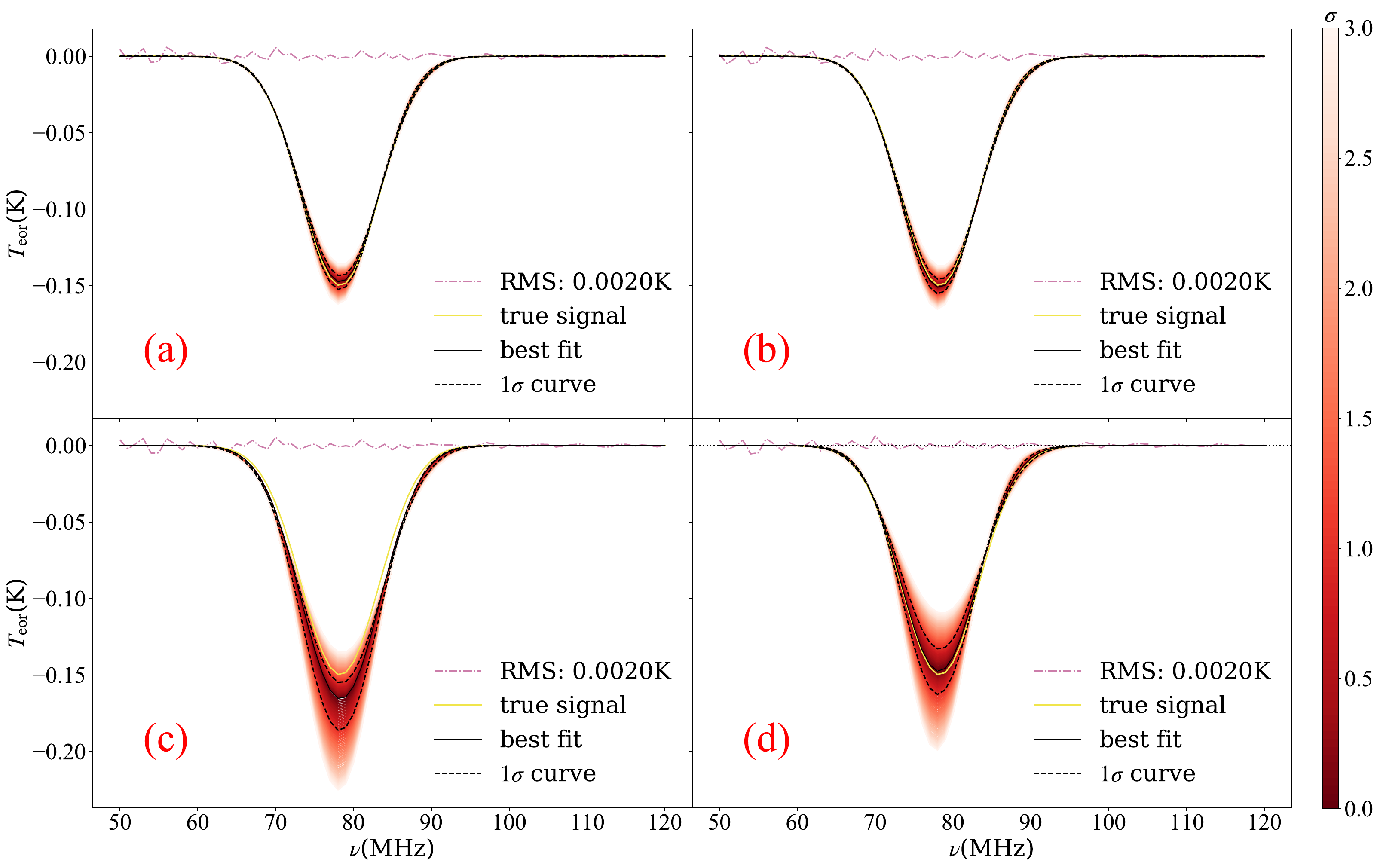}
		\caption{Fitting results for \textbf{Gaussian model} with different antennas (optimized blade antennas with $W=1.25, 1.5, 2$\,m and bow-tie blade antenna with $L=3$\,m). The reddish-purple dash-dotted line represents the residuals after fitting and removing both foreground and absorption features. The yellow solid line denotes the input absorption feature, while the black solid is the recovered best-fit line. Also, we plot dozens of lines to represent different error levels with different shades of colour (see the colour bar to the right of the figure). For clarity, we intentionally plot two 1$\sigma$ lines in (dashed black lines).}
		\label{fig:11_extreme_recovered}
	\end{figure*}

	Fig~\ref{fig:11_extreme_recovered} shows the fitting results for \textbf{Gaussian model} based on the four extreme frequency-dependent antennas. panels (a) and (b) still show near-perfect signal reconstruction results. The reconstructed EoR absorption feature in panel (c) deviates obviously from the input feature and the ranges of the parameter distribution of both panel (c) and (d) are significantly wider. Fig~\ref{fig:11_extreme_recovered} shows that, assuming the beam is accurately known, VZOP still performs well even with extreme frequency dependence.

	\subsection{Number of Declination Bins}
	It can be seen from Fig.~\ref{fig:11_recovered} and Fig.~\ref{fig:22_recovered} that our reconstructed EoR absorption feature and input feature are highly coincident for all antennas when we sample 10 declination bins, whether it is \textbf{Gaussian model} or \textbf{EDGES model}. These results show that VZOP fitting reduces the effects of antenna chromaticity very well.
	
\begin{figure}
		\centering
		\begin{subfigure}{\columnwidth}
			\centering
			\includegraphics[width=0.9\linewidth]{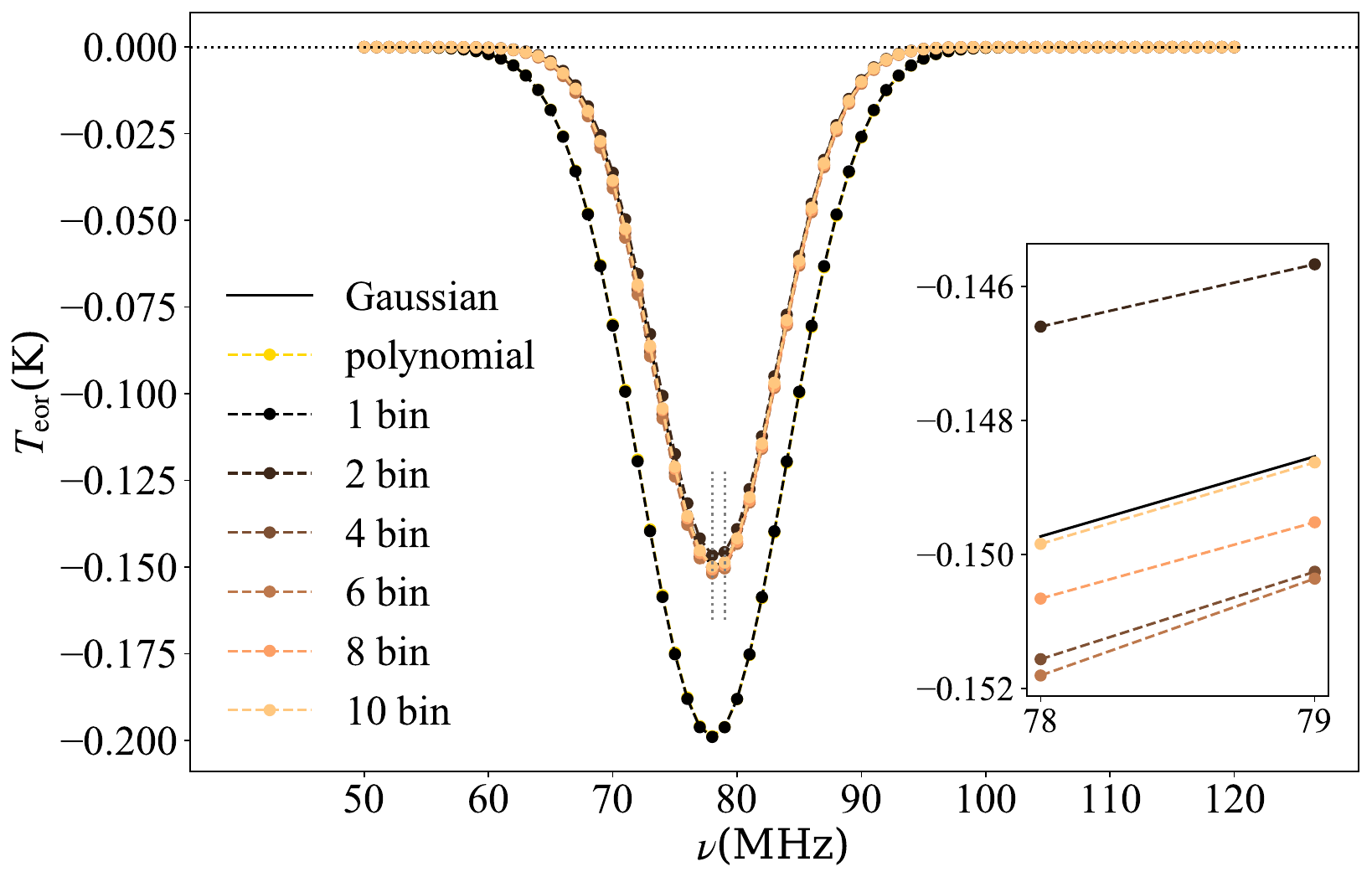}
		\end{subfigure}
		\begin{subfigure}{\columnwidth}
			\centering
			\includegraphics[width=0.9\linewidth]{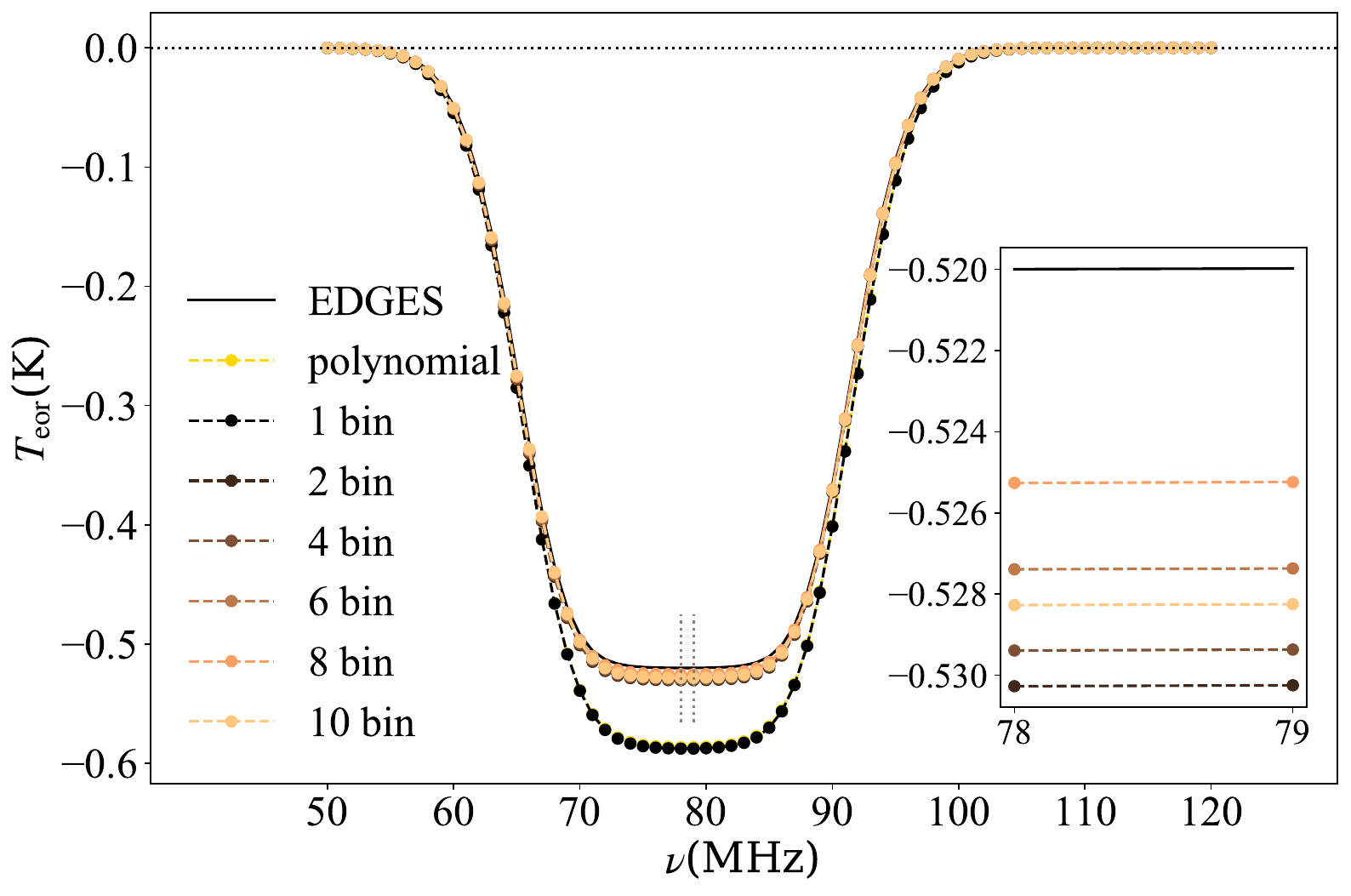}
		\end{subfigure}
		\caption{Fitting results using different numbers of declination bins for \textbf{Gaussian model} (top panel) and \textbf{EDGES model} (bottom panel) with optimized blade antenna.}
		\label{fig:different_bins}
	\end{figure}

	To determine at least how many bins obtain the best results, we refit with different $N_\mathrm{bin}$ for two models. We use an optimized blade antenna as the fiducial antenna because of the weakest frequency correlation. Fig~\ref{fig:different_bins} shows the fitting results, where the common polynomial algorithm is added for comparison.

	When we use only 1 bin, VZOP fitting will degrade into the polynomial fitting in theory. As can be seen from the two panels in Fig~\ref{fig:different_bins}, as expected, the result obtained using 1 bin is highly coincident with that obtained by the common polynomial fit algorithm. However, with only 2 bins, the reconstructed absorption feature is almost identical to the input feature for both cases. There is no significant improvement in fitting even if continually increasing $N_\mathrm{bin}$. The results in Fig.~\ref{fig:different_bins} show that for the optimized blade antenna, only 2 bins are sufficient to recover the 21\,cm absorption feature. For a well-designed antenna, VZOP will carry enough information of the beam with only a few declination bins. 
	
	\section{Inaccurate Antenna Beam Pattern}
	\label{sec:IABP}
	We have seen that the VZOP fits very well for a wide range of commonly used antennas. The above results confirm our assumption, that as long as the information of the antenna beam is brought into the fitting by some method, the foreground can be deducted efficiently. Therefore, VZOP fits very well for a wide range of commonly used antennas when the beam can be accurately known, and usually can extract 21\,cm absorption feature by using just 2 declination bins. Unfortunately, in actual observations, the antenna beam pattern can hardly be precisely measured due to various systematics such as reflections from surrounding objects, ionospheric effects, noise inside the instrument, etc. Even performing a systematic correction, it is expected to still have 1\% - 10\% residual systematics. Therefore, in this section, we will study the performance of VZOP when there are errors in the beam. Specifically, we will use an accurate beam in simulation and add a Gaussian error to each pixel of the beam model used in VZOP. Note that the random error will be smoothed out in the process of arithmetic averaging, the error of the 24-hour averaged beam will be much less than that of the origin antenna beam. 
	
	The errors at different pixels and frequencies are not completely random, as they arise from systematics. In other words, the errors at different pixels and frequencies exhibit correlation. Therefore, we assume that the errors at different pixels are random, but the errors at different frequencies within the same pixel are correlated. We studied four models: constant, linearity, quadratic and cosine. The details of them are shown in Table~\ref{tab:four_error_models}. We consider these four error models and simulate observations using three fiducial antennas respectively (using the \textbf{Gaussian model}). For all the situations, using the VZOP considering 10 declination bins can successfully extract the 21\,cm absorption feature, even if the error reaches 10\% which is a value the current correction technology can completely make the remaining error of the beam less than it.
	\begin{table}
		\centering
		\caption{Functions of four beam error models. Where $\mathrm{e}_B$ is the relative error between measured and real value,  $\mathbf{n_0}$ is any fixed position and $\nu_0=50$\,MHz. Unit of frequency $\nu$ in each function is MHz.}
		\label{tab:four_error_models}
		\begin{tabular}{|l|l|}
			\toprule
			Model & Function\\ \hline
			constant & $\mathrm{e}_B(\nu,\mathbf{n_0})=\mathrm{e}_B(\nu_0,\mathbf{n_0})$\\ \hline
			linearity &	$\mathrm{e}_B(\nu,\mathbf{n_0})=\left(\dfrac{\nu-85}{35}\right)\times\mathrm{e}_B(\nu_0,\mathbf{n_0})$\\ \hline
			quadratic & $\mathrm{e}_B(\nu,\mathbf{n_0})=\left[2\left(\dfrac{\nu-85}{35}\right)^2-1\right]\times\mathrm{e}_B(\nu_0,\mathbf{n_0})$\\ \hline
			cosine & $\mathrm{e}_B(\nu,\mathbf{n_0})=\cos{\left(\dfrac{2\pi}{10}\nu\right)}\times\mathrm{e}_B(\nu_0,\mathbf{n_0})$\\
			\bottomrule
		\end{tabular}
	\end{table}
	
\begin{figure*}
		\centering
		\includegraphics[width=\linewidth]{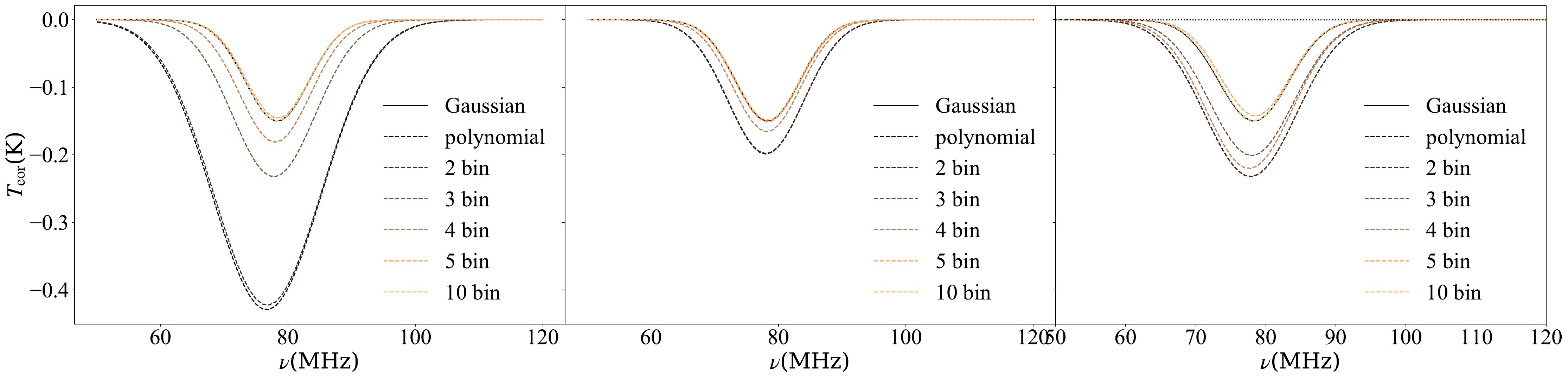}
		\caption{Fitting results for \textbf{Gaussian model} using different number of declination bins with nominal blade (left), optimized blade (middle) and bow-tie blade antenna (right). For comparison, we also plot the results of the polynomial fit.}
		\label{fig:11_different_bins_cosine0.1}
	\end{figure*}

\begin{figure*}
		\centering
		\includegraphics[width=\linewidth]{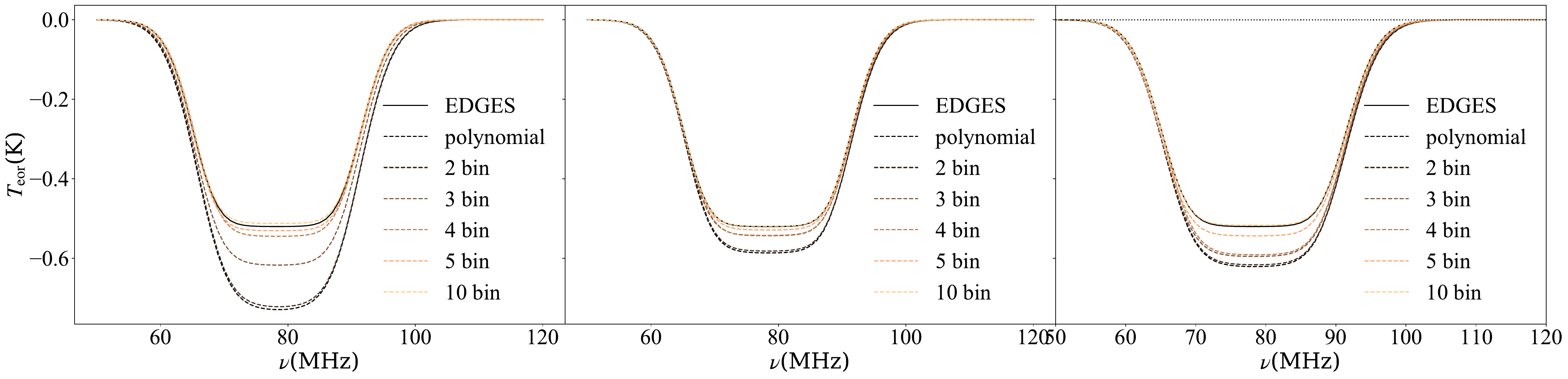}
		\caption{Fitting results for \textbf{EDGES model} using different number of declination bins with nominal blade (left), optimized blade (middle) and bow-tie blade antenna (right). For comparison, we also plot the results of the polynomial fit.}
		\label{fig:22_different_bins_cosine0.1}
	\end{figure*}

	Compared with polynomial fitting, VZOP has a higher degree of freedom, so VZOP performs well even with a relative error of up to 10\%. Only considering the cosine error model with 10\% error, we show the effect of the number of declination bins on the fitting results under the three fiducial antennas in Fig.~\ref{fig:11_different_bins_cosine0.1} (\textbf{Gaussian model}) and Fig.~\ref{fig:22_different_bins_cosine0.1} (\textbf{EDGES model}). From the six panels in the two figures, on the whole, when only two declination bins are used, the fitting results are close to that of the polynomial fitting. As the number of bins increases, the fitting results gradually become better, and the output absorption feature mostly coincides with the input feature when 5 bins are used. But there are some exceptions. Such as the middle panel of the Fig.~\ref{fig:11_different_bins_cosine0.1} shows an abnormal situation that the output feature of $N_\mathrm{bin}=3$ is more accurate than $N_\mathrm{bin}=4$. This may just be due to the error in our model. When $N_\mathrm{bin}=3$, the error is offset by a large part. When the number of bins continues to increase, the error increases again. On the other hand, comparing different columns of the two figures (i.e. different antennas), using 10 bins gives an exact fit for any antenna. When the number of bins is too small to extract accurately the input absorption feature, the fitting results of the optimized blade antenna that is weaker frequency dependence are better.
	
\begin{figure}
		\centering
		\begin{subfigure}{\columnwidth}
			\centering
			\includegraphics[width=0.9\linewidth]{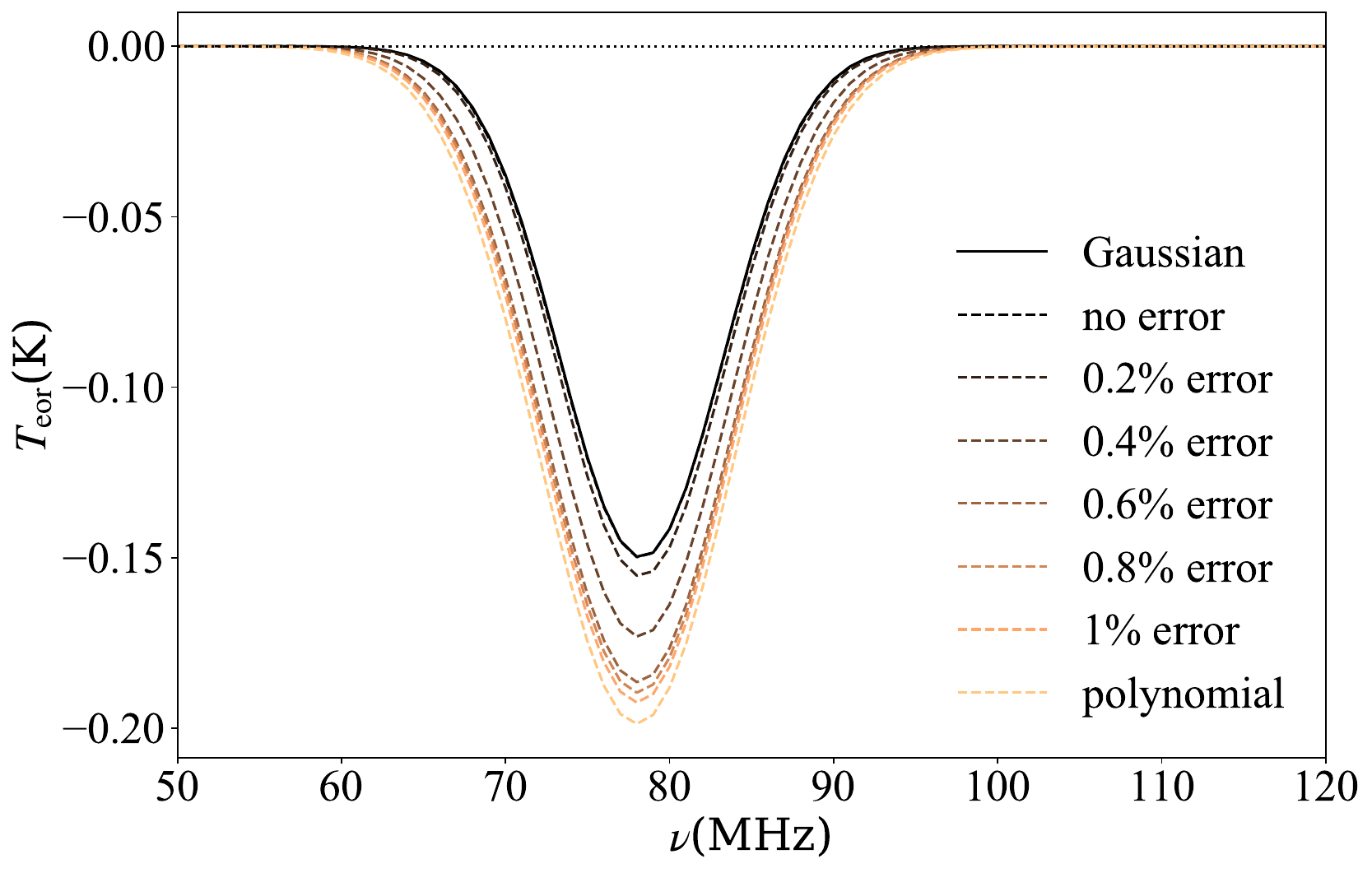}
		\end{subfigure}
		\begin{subfigure}{\columnwidth}
			\centering
			\includegraphics[width=0.9\linewidth]{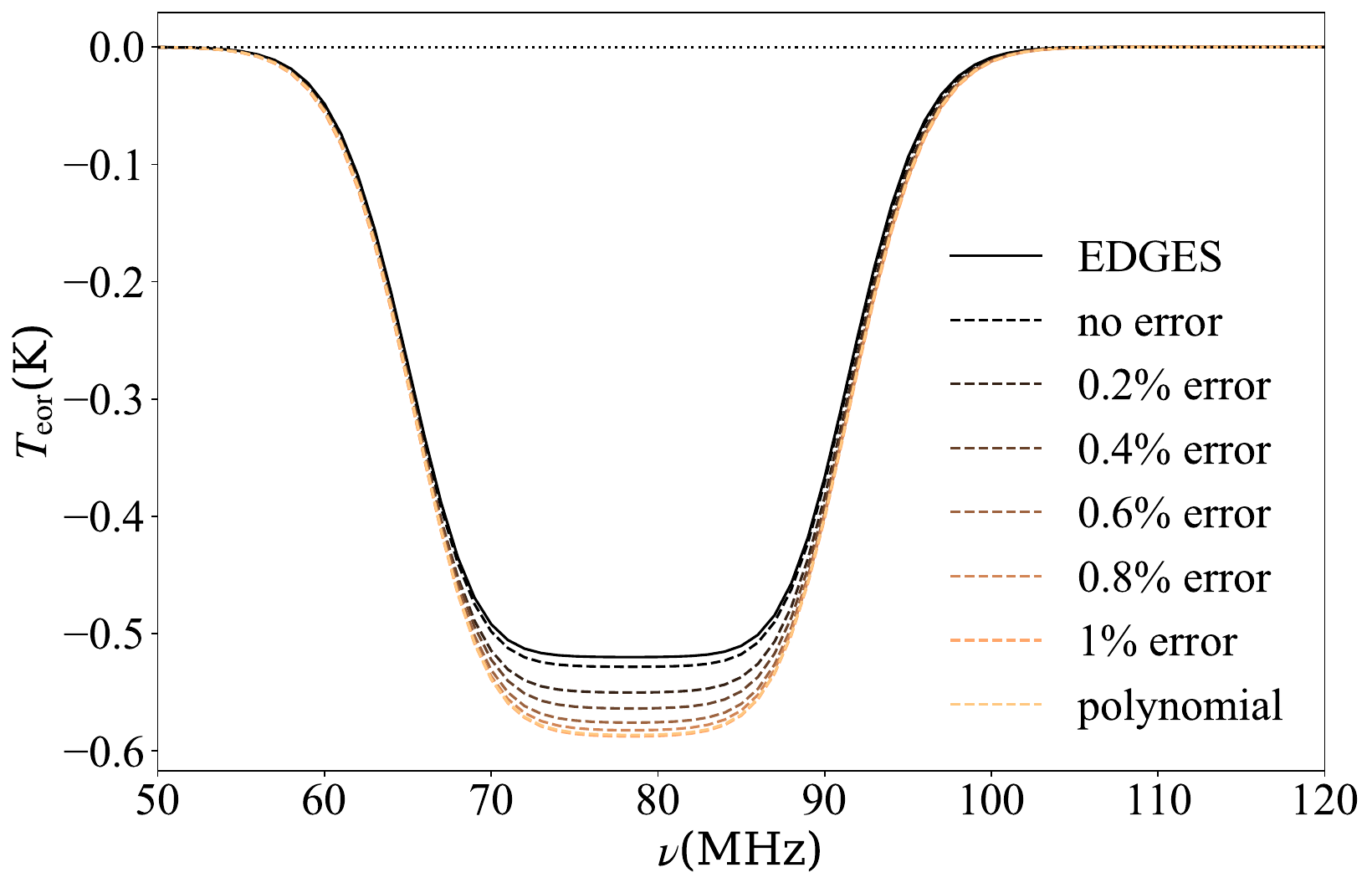}
		\end{subfigure}
		\caption{Fitting results when adding different Gaussian errors to the optimized blade antenna for \textbf{Gaussian model} (top panel) and \textbf{EDGES model} (bottom panel) assuming the errors are completely random. We also plot the results from polynomial fitting for comparison.}
		\label{fig:all_random}
	\end{figure}

	Based on the aforementioned results, VZOP works well even in the presence of 10\% errors. Furthermore, we evaluate the performance of VZOP under the worst-case scenario, where the errors at different pixels and different frequencies are completely random. We added different Gaussian errors to the optimized blade antenna and re-extracted the 21\,cm absorption feature, and the signal reconstruction results are shown in Fig.~\ref{fig:all_random}. The output absorption features are highly consistent with the input features for both two signal models when there are no errors. As the error increases, the reconstruction results gradually become worse and closely approximate to that of polynomial fitting when the error magnitude reaches 0.6\%. In other words, 0.6\% error is enough to cause VZOP to lose its advantage. However, it is worth noting that as the error continues to increase, the fitting performance of VZOP does not deteriorate further. After testing, no matter how large the error is, VZOP will not be worse than the polynomial fitting, because VZOP has more degrees of freedom. We have even set the gain on each pixel to an arbitrary value, and the results did not become worse than that of polynomial fitting. Hence, on the one hand, the actual errors of the antenna beam are certainly not entirely random. On the other hand, even in the worst-case scenario, the fitting results of VZOP do not deteriorate compared to that of common polynomial fitting.

    \section{Discussions and Conclusions}
    \label{sec:Discussions and Conclusions}
    In this work, we reconstruct the cosmic 21\,cm absorption feature from foreground contamination using VZOP - an improved polynomial fitting algorithm - by defining a 24-hour averaged beam model. We plan to apply VZOP in the forthcoming LACE. Nevertheless, in its capacity as a general algorithm, VZOP can be employed in any single antenna-based global 21\,cm signal experiment. VZOP is introduced to address a specific problem that the chromaticity of the antenna couples the spatial structure of the Galactic bright temperature map into the frequency structure, making the common polynomial fitting work poorly and hindering the reconstruction of the cosmic 21\,cm signal. If we know the antenna beam, we can invert the true global spectrum, which is the idea utilized by VZOP that brings the information of the beam into the reconstruction model such as polynomial fitting. From a mathematical viewpoint, VZOP is equivalent to an improved "polynomial fitting" model, in which the zeroth-order item is a quantity that varies with frequency and beam-related information is included in the zeroth-order item. The spectral structure introduced by the antenna chromaticity can be deducted by using the beam information to fine-tune the polynomial "constant term".

    Assuming the antenna beam can be accurately measured, VZOP will accurately deduct the extra structure introduced by chromaticity (even with only 2 declination bins), and the reconstructed 21\,cm absorption feature is highly coincident with the input feature, which is consistent with expectations. The fact that VZOP works with only 2 bins suggests that additional structures may be described with only one parameter. In addition, VZOP performs well even for extremely frequency-dependent antennas.

    In actual observation, we cannot accurately measure the antenna beam due to various system errors. It is not realistic to assume an accurate antenna beam pattern in the solving process. Therefore, taking into account that the errors are from systematics, such as the antenna dimensions errors, we assume that they are random at different pixels and follow several models outlined in Table.~\ref{tab:four_error_models} at different frequencies. VZOP can still successfully extract the 21\,cm absorption feature even if the error reaches 10\%. However, in the presence of errors, more declination bins may be required to ensure that VZOP performs well. For example, under the cosine error model, at least 5 bins are required to recover the \textbf{Gaussian model} and the \textbf{EDGES model}.

    Finally, we investigate the fitting results of VZOP under the worst-case scenario, where the errors are completely random across different pixels and frequencies. As long as the error exceeds 0.6\%, it is enough to interfere with VZOP's identification of additional structures and make VZOP lose its advantage. However, an interesting phenomenon is that no matter how large the error is, the fitting results of the VZOP are not worse than that of the polynomial fit. The reason is that the essence of VZOP is to fine-tune the polynomial "constant term" to identify the extra structure introduced by chromaticity, increasing the error only reduces the ability to identify the extra structure. Based on empirical observations, it can be seen that as the error increases, the variation of the zeroth-order term to frequency becomes progressively smaller, eventually converging to a "constant". This feature ensures that even in the worst-case scenario, the performance of VZOP will not be worse than the common polynomial fitting, let alone the fact that the errors are certainly not completely random.

    In summary, the fitting performance of VZOP depends on the accuracy of beam measurements and the error structure. More bins are required to recover the 21\,cm absorption feature when errors exist due to the error reducing the amount of information carried by each bin, which is why VZOP's performance deteriorates in the case of completely random errors. When the error is completely random, the degree of freedom of the error becomes too high, resulting in little information carried by each bin. It is worth noting that using too many bins can lead to overfitting. To address this issue, using a narrower frequency bandwidth could be considered. Because then there will be more frequency channels, which will enable the use of more declination bins and improve the performance of VZOP in the presence of beam errors. This could be an area of future research.

    In the simulation, VZOP can successfully extract the cosmic 21\,cm absorption feature, but the real observation may be more complicated. Whether it is the \textbf{Gaussian model} or the \textbf{EDGES model}, it is only the approximate description of the real cosmic 21\,cm signal. On the other hand, the true residual error may not be described by a few simple models. In the future, we will continue to study the performance of VZOP in more realistic observations.
	
	\section*{Acknowledgements}
	
	TL, JG and QG acknowledge the support from the Ministry of Science and Technology of China (grant No. 2020SKA0110200). HYS, QZ and JW acknowledge the support from the Ministry of Science and Technology of China (grant No. 2020SKA0110100). HYS acknowledges the support from the National Science Foundation of China (11973070), the Key Research Program of Frontier Sciences, CAS, Grant No. ZDBS-LY-7013 and the Program of Shanghai Academic/Technology Research Leader. QZ acknowledges the support from the National Science Foundation of China (11973069). Finally, we are sincerely grateful to the referee for constructive and valuable suggestions and comments, which help us improve the manuscript. 
	%%%%%%%%%%%%%%%%%%%%%%%%%%%%%%%%%%%%%%%%%%%%%%%%%%
	\section*{Data Availability}

	The code and data underlying this article will be shared on reasonable request to the corresponding authors.

	
	
	\bibliographystyle{mnras}
	\bibliography{reference} % if your bibtex file is called example.bib

	%\begin{thebibliography}{99}
	%\end{thebibliography}
	
	%%%%%%%%%%%%%%%%%%%%%%%%%%%%%%%%%%%%%%%%%%%%%%%%%%

	% Don't change these lines
	\bsp	% typesetting comment
	\label{lastpage}
\end{document}